\pdfoutput=1
\documentclass[11pt,a4paper]{article}

\usepackage[utf8]{inputenc}
\usepackage[T1]{fontenc}

\usepackage[margin=1in]{geometry}
\usepackage{graphicx}
\DeclareGraphicsExtensions{.png,.pdf,.jpg}
\usepackage{amsmath,amssymb,amsfonts}
\usepackage{mathtools}
\usepackage{bm}
\usepackage{xcolor}
\usepackage{authblk}
\usepackage{hyperref}
\usepackage{url}
\usepackage{caption}
\usepackage{longtable}
\usepackage{booktabs}
\usepackage{array}

\hypersetup{
    colorlinks=true,
    linkcolor=blue,
    urlcolor=blue,
    citecolor=blue
}

\title{A Quantum-Inspired Two-Dimensional Lattice Polarized Radiative Transfer Simulator}

\author[1]{Shan Zeng\thanks{Correspondence: \href{mailto:shan.zeng@hotmail.com}{shan.zeng@hotmail.com}}}
\author[2]{Bing Lin}
\author[2]{Ali Omar}
\affil[1]{Eleven Dimensions, LLC, Yorktown, VA, 23693, USA}
\affil[2]{NASA Langley Research Center, Hampton, VA, 23681, USA}

\date{}

\begin{document}

\maketitle

\begin{abstract}
\noindent
A quantum inspired framework was developed for polarized radiative transfer in the atmosphere to jointly model multiple scattering, absorption, and emission processes. The radiative field is represented by the Stokes vector and discretized using an n-directional two-dimensional spatial lattice on 3-dimensional uniformly distributed zenith and azimuth angular vectors. The formulation of this quantum computation system is based on the Lattice Boltzmann Method, which incorporates a physics-based polarized phase matrix. A classical lattice solver employing identical spatial and angular discretization and physical modeling as in quantum algorithm is implemented. The corresponding quantum algorithm is evaluated using the IBM Qiskit quantum simulator, enabling direct comparison of numerical accuracy to a classical simulation reference. The results demonstrate that fully physics-based polarized radiative transfer can be embedded within a quantum lattice framework, which provides an initial benchmark toward scalable quantum atmospheric radiative transfer and future quantum climate modeling.

\end{abstract}

\vspace{1em}

\section{Introduction}

The Atmospheric radiative transfer plays a central role in weather prediction and climate modeling, governing the radiative interaction of atmospheric gases, aerosols, and clouds through absorption, emission, and scattering [1]. Accurate treatment of these processes numerically is essential for simulating energy balance, remote sensing, and climate feedback modelling. Operational models such as MODTRAN have been widely used to calculate atmospheric transmission and radiance for atmospheric correction and spectral-image analysis [2], while the Atmospheric and Environmental Research suite, including LBLRTM and RRTM, supports applications in remote sensing, numerical weather prediction, and climate modeling [3]. In particular, multiple scattering and polarization effects are increasingly important in these numerical calculations for high-fidelity simulations and interpretation of modern satellite and ground-based measurements. However, solving the polarized radiative transfer equations in realistic geometries remains computationally demanding, especially when spectral resolution, complex phase matrices, and large spatial domains are required [1, 4].

Classical radiative transfer solvers typically rely on either deterministic methods such as discrete ordinates [4] and spherical harmonics [5], or stochastic methods such as Monte Carlo techniques [1, 6]. While these approaches are well established, their computational cost increases rapidly with the number of angular directions, spectral channels, and spatial dimensions. This scaling issue becomes a major bottleneck when coupling radiative transfer with dynamical cores in general circulation models or when resolving polarization and/or hyperspectral and directionally dependent scattering in three-dimensional heterogeneous atmospheres

Although machine-learning approaches have recently been proposed to accelerate both forward radiative-transfer calculations and inverse (retrieval) simulations by learning fast surrogate models from precomputed datasets, their performance fundamentally relies on the availability of large volumes of high-fidelity numerical simulations for training and validation [7-8]. Consequently, meaningful further acceleration of large-scale, high-fidelity radiative-transfer simulations is expected to rely primarily on advances in high-performance computing architectures and, in the longer term, on emerging quantum-computing paradigms for scientific simulation [9]. Quantum-lattice formulations provide a natural framework [10] for representing transport and local interactions through unitary operations. Related quantum-walk algorithms have also been proposed for Monte Carlo particle transport and photon-interaction simulations [11-12]. These ideas have led to quantum radiative transfer models that mirror the structure of classical radiative transfer methods, while offering a pathway toward future hardware acceleration on quantum devices.

Among numerical schemes for the radiative transfer equation, we adopt the lattice-Boltzmann framework because it aligns naturally with quantum circuits and scales efficiently in both spatial and angular resolution. Although quantum-lattice methods for fluid and particle transport have been studied for several years, their application to atmospheric radiative transfer and remote sensing remains limited, with existing studies largely confined to theoretical formulations and proof-of-concept demonstrations [10-12]. In this work, we introduce a lattice-based quantum algorithm of atmospheric radiative transfer that explicitly incorporates multiple scattering, absorption, emission, and polarization and scale it up to n-directional and 2-dimensional lattices. The radiative field is represented in terms of the Stokes vector (I, Q, U, V), and scattering interactions are modeled through a physics-based phase-matrix formulation. A discrete velocity lattice is employed to describe angular transport, enabling the propagation and scattering processes of radiative transfer to be mapped to unitary operations in a quantum circuit. As a concrete realization, we construct and examine an n-direction two-dimensional (D2Q9) lattice and implement Rayleigh scattering within a polarized quantum radiative transfer framework. Though Rayleigh scattering is mainly used for small atmospheric particles, the phase matrices used could be easily adapted to other phase matrices such as those for aerosol, cloud and precipitation-sized hydrometeor droplets. Thus, the algorithm developed here builds a quantum computing foundation in solving general radiative transfer processes of atmospheric gases, aerosols, clouds, precipitation, etc.

The proposed approach preserves the essential physical structure of classical lattice radiative transfer methods while providing a direct mapping to quantum operations. This makes it possible to study, in a controlled setting, how polarized radiative transfer processes can be embedded into quantum algorithms. At the same time, a classical reference solver is constructed using the same lattice and physical assumptions, enabling a direct comparison and validation between classical and quantum formulations.

This study aims to establish a foundational framework in quantum polarized radiative transfer modeling for atmospheric and climatic sciences. By demonstrating the treatment of polarization, scattering, absorption, and emission on a 2-dimensional lattice using quantum operations, we provide the first step toward more comprehensive quantum atmospheric solvers. Future extensions will incorporate spectrally resolved molecular absorption, aerosol scattering models, and more general cloud scattering matrices, as well as the extension from two-dimensional lattices to fully three-dimensional lattice geometries. Ultimately, this line of research seeks to establish physics-based quantum radiative transfer modeling tools.

The quantum radiative transfer model and its circuit implementation are presented in Section 2. Numerical results and validation against the classical reference are given in Section 3. Section 4 provides the conclusion and an outlook on future work.

\section{Quantum Radiative Transfer Model}

Radiative transfer is a physical process describing the transport of electromagnetic wave (EMW) and energy through participating medium accounting for absorption, scattering, emission, and polarization processes [13]. In the atmosphere, radiation propagation is governed by the combination of these effects and strongly influenced by interactions among EMWs through atmospheric agents. Radiative transfer equation (RTE) provides a general mathematical description of these physical processes as summarized in Eq (1).

\begin{equation}
\begin{split}
\frac{1}{c}\frac{\partial S(r,\widehat{\Omega},t)}{\partial t} + \widehat{\Omega}\cdot\nabla S(r,\widehat{\Omega},t) 
={} & -k_{ext}(r)\,S(r,\widehat{\Omega},t) \\
    & + \int_{0}^{4\pi} k_{scat}(r)\,P(\widehat{\Omega},\widehat{\Omega}^{'})\,S(r,\widehat{\Omega}^{'},t)\,d\Omega^{'} + J(r,\widehat{\Omega},t)
\end{split}
\tag{1}
\end{equation}

Where S ([I, Q, U, V]\textsuperscript{T}) is the Stokes vector. c is the speed of light in the medium. $\widehat{\Omega} \bullet \nabla S$ is the directional derivative which is equal to ${\ \Omega}_{x}\frac{\partial S}{\partial x} + {\ \Omega}_{y}\frac{\partial S}{\partial y} + \ {\ \Omega}_{z}\frac{\partial S}{\partial z}$. $k_{ext}$ is the extinction coefficient and equal to the sum of the scattering coefficient ($k_{scat}$) and absorption coefficient ($k_{abs}$). $P(\widehat{\Omega},\ {\widehat{\Omega}}^{'})$ is the polarized scattering phase matrix. $\widehat{\Omega}$ represents ray directions, and \emph{J} denotes an emission or source term.

A wide range of numerical methods has been developed to solve the RTE. In this work, we investigate lattice-based numerical approaches that are particularly well aligned with quantum-computing formulations. Specifically, we employ a lattice Boltzmann--based radiative transfer framework belonging to the class of discrete-ordinates methods. The spatial domain r is discretized on a two-dimensional lattice using a D2Q9 scheme, while the angular domain $\widehat{\Omega}$ is discretized independently through a three-dimensional directional representation parameterized by uniformly distributed zenith and azimuth angles. Such unified discretization technique enables consistent modeling of polarized radiative transfer and facilitates direct mapping of RTE to both classical and quantum lattice algorithms, readily applicable in atmospheric modelling and realistic satellite and other remote-sensing applications.

\subsection{Boltzmann Lattice Method}

The lattice method is naturally aligned with quantum computing because its update rules are composed of local, structured, and reversible operations that can be directly mapped to quantum circuits [14]. A simple intensity RTE with isotropic scattering in the upward and downward two directions has been demonstrated to be solvable [10] using a one-dimensional two-velocity (D1Q2) lattice model, as illustrated in Fig.\,1(a). This study aims to address the general solution of the RTE in a realistic atmosphere, including the combined effects of absorption, emission, scattering, polarization, and external radiative sources. In this work, radiation transport is simulated using a D2Q9 lattice model (Fig.\,1b). The lattice comprises 32 grid points in the one-dimensional D1Q2 case and a $16 \times 16$ grid for the two-dimensional D2Q9 case.

\begin{figure}[htbp]
\centering
\includegraphics[width=4.81111in,height=2.35015in]{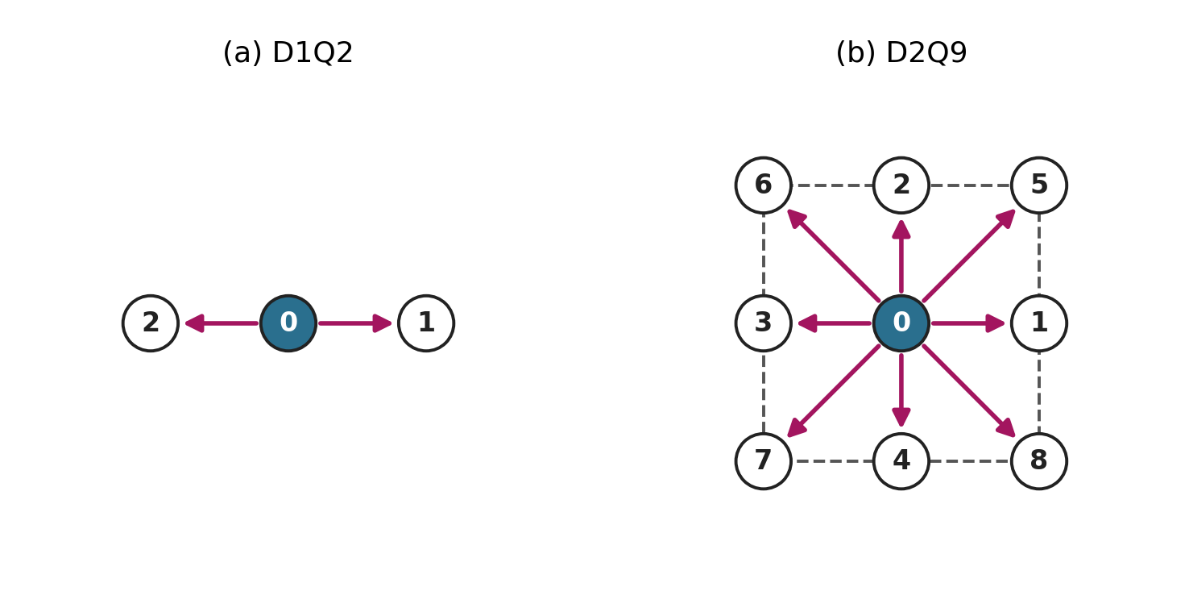}
\caption{Fig. 1. Schematics of D1Q2 (a) and D2Q9 (b) lattices.}
\end{figure}

\subsubsection{Radiative Transfer Equation in Lattice}

The D1Q2 lattice-based formulation of the transfer equation for fluid dynamics has been extensively studied in the literature. However, existing approaches for EMWs are limited to the scalar radiative intensity (Igarashi et al., 2024) and cannot be directly extended to the vector (polarized) RTE formulated in terms of the full Stokes parameters [I, Q, U, V]. Building on the basic scalar RTE solution, we explore the lattice representation from D1Q2 to a two-dimensional D2Q9 stencil, yielding the discrete update rule given in Eq. (2), where full polarization has been represented.

The D2Q9 stencil provides a minimal and computationally convenient mapping from angular directions onto spatial streaming, but it restricts propagation to eight lattice directions, namely lattice streaming along (-1, 0, 1) for both x and y axis, as illustrated in Figure 2.

An angular resolution of 32--64 directions provides a good balance between physical accuracy and computational feasibility. In the quantum implementation, however, the collision operator must be embedded in a larger unitary operator acting on an extended Hilbert space. The number of qubits required for a direction register grows logarithmically with the number of directions. Because the implementation uses separate registers for the incoming and outgoing directions, doubling the number of directions adds one qubit to each register, increasing the statevector memory requirement by approximately a factor of four. Quadrupling the number of directions adds two qubits to each register and increases the memory requirement by approximately a factor of sixteen, assuming that all other registers remain unchanged.

For this demonstration, we therefore use 16 directions. The method can be extended to more directions when additional GPU memory is available. Importantly, the lattice Boltzmann method maps naturally to a quantum-computing implementation because it separates the dynamics into direction-controlled streaming operations acting on the lattice registers and local scattering and source updates applied at each lattice site.

\begin{equation}
\begin{split}
S_{d}\!\left(x + c_{d,x}\Delta t,\, y + c_{d,y}\Delta t,\, t + \Delta t\right) - S_{d}(x,y,t) 
={} & -k_{ext}\,\Delta t\,S_{d}(x,y,t) \\
    & + k_{scat}\,\Delta t\sum_{d^{'}=0}^{n_{d}-1}\omega_{d^{'}}P(d,d^{'})\,S_{d^{'}}(x,y,t) \\
    & + \Delta t\,J_{d}(x,y,t)
\end{split}
\tag{2}
\end{equation}
 where \emph{S\textsubscript{d}} is a discretized Stokes vector (I, Q, U, V). P is phase function weighted by the angular quadrature weights, $\omega_{d},\ d = 0,\ 1,\ \ 2,\ldots\ ,8$, that are equal to 1/8 for 8 directions and 0 for the rest point, d means discrete income direction index while $d^{'}$is for outgoing direction index. x and y are the lattice index in the 2D x and y plane.

We adopt the lattice-update strategy of Igarashi et al. (2024) at each lattice site. In each time step, the radiative field of the Stokes vector is first amplitude-encoded into the quantum register. The local collision step is then applied in sequence, consisting of the scattering and depolarization, absorption/extinction, and emission (source injection) processes. After these local updates, a streaming operation propagates the Stokes Vector along the lattice according to the pair of incoming-outgoing directions to neighboring lattice sites. Finally, the quantum state is measured and decoded to classical values to recover the updated radiative field.

For small spheric particles, to keep the scattering term to be physically consistent, the phase function should satisfy the discrete normalization ($\sum_{d^{'} = 0}^{n_{d} - 1}{\omega_{d^{'}}P\left( d,d^{'} \right)} = 1$) [15], i.e., $\omega_{d^{'}} = \frac{4\pi}{n_{d}}$.

\subsubsection{Encoding and Quantum State Initiation}

For quantum RT calculations, the first step is to transfer classical RT values to quantum qubits. Compared with intensity-only transport, a Stokes register of two qubits is incorporated for polarization needs:

$|00\rangle \equiv I$, $|01\rangle \equiv Q$, $|10\rangle \equiv U$, $|11\rangle \equiv V$ (3)

The discretized, polarization field is stored as a real-valued Stokes Vector of stream and source $P(st,sl,d_{in},\ d_{out},y,x) \in R$, where st indexes Stokes components, sl denotes source or live stream index, $d_{in}$ and $d_{out}$ the propagation direction index, and (y, x) the spatial lattice index. In our simulation, we use $n_{st} = 4$, $n_{sl} = 2$, $n_{d} = n_{din} = \ n_{dout} = 8$, $n_{x} = n_{y} = M = 16$. This tensor is mapped to a register-ordered quantum state $|\phi(x,y,d_{in},\ d_{out},sl,st)\rangle$ (see, Eq. 4) via amplitude embedding (Shuld and Petruccione, 2019), which encodes a classical vector of 2n into the amplitudes of a quantum state of n qubits. The input vector is $\mathcal{l}_{2}$-normalized so that it defines a valid quantum state that can be prepared by a unitary state-preparation circuit on a quantum computer. Explicitly,

\begin{equation}
\left| \phi_{0} \right\rangle = \frac{1}{\left\|\phi_{0} \right\|}\sum_{i_{st},i_{sl},d_{in},d_{out},i_{y},i_{x}}^{}\phi_{in;i_{st},i_{sl},d_{in},d_{out},i_{y},i_{x}}\,\left| i_{st} \right\rangle_{st}\left| i_{sl} \right\rangle_{sl}\left| d_{in} \right\rangle_{in}\left| d_{out} \right\rangle_{out}\left| i_{y} \right\rangle_{y}\left| i_{x} \right\rangle_{x}
\tag{4}
\end{equation}

where the $d_{in}$ and $d_{out}$ register encodes the angular direction using ${log}_{2}N_{d}$ qubits, with directions equally distributed over the zenith and azimuth angles around the globe. The x and y registers encode the lattice indices along the East-West and North-South axes. For sl register $|0\rangle_{sl}\text{ }$ represented live transported intensity and $|1\rangle_{sl}$ means emission/source term. The subscript "in" of the quantum state $|\phi_{in}\rangle$ denotes the input matrix, while "0" ($|\phi_{0}\rangle$) denotes the quantum state after encoding.

We initialize the source term in two different experiments. In the first experiment, we follow the same initialization procedure as in Igarashi et\,al. (2024). Let $S$ denote the live stream amplitude vector encoded in the $|0\rangle_{s}$ subspace and let $J$ denote the external emission source amplitude vector encoded in the $|1\rangle_{s}$subspace. The live stream amplitudes are initialized to be zero for all lattice points and propagation directions, i.e., $S = 0$. The source term is initialized within a spatial region covering $\pm 25\%$ of the domain, centered at the origin, with a uniform directional emission amplitude whose sum equals 1 across all propagation directions.

In the second experiment, the source region is restricted to the central lattice points with gaussian distribution only to examine longer-range transport patterns in both intensity and polarization. All other settings, including the emission amplitude (J=1) and scattering ($k_{scat}$= 0.5) and extinction ($k_{ext} = 2.5$) coefficients, are kept the same as those used in Igarashi et\,al. (2024). Note also photons that transfer out of the domain are fully absorbed and do not return (absorbing boundaries).

The following sections focus on the treatment of radiative transfer processes, describing light--atmospheric agent interactions implemented using quantum operations. These processes are applied sequentially and represented as a product of operators ($O_{\text{emis}}$, $O_{\text{scat}}$, $O_{\text{abs}}$, and $O_{\text{propa}}$) as shown in Eq.\,(5) and Figure 2:

\begin{equation}
\left| \phi_{out} \right\rangle = {|0\ldots 0\rangle }_{a}\ \frac{1}{\left\| O_{emis} \right\| \times \left\| O_{scat} \right\| \times \left\| O_{abs} \right\| \times \left\| O_{propa} \right\| \times \left\| \phi_{0} \right\|}(O_{emis} \cdot O_{scat} \cdot O_{abs} \cdot O_{propa})\left| \phi_{in} \right\rangle\ 
\tag{5}
\end{equation}

\begin{figure}[htbp]
\centering
\includegraphics[width=5.46875in,height=1.91804in]{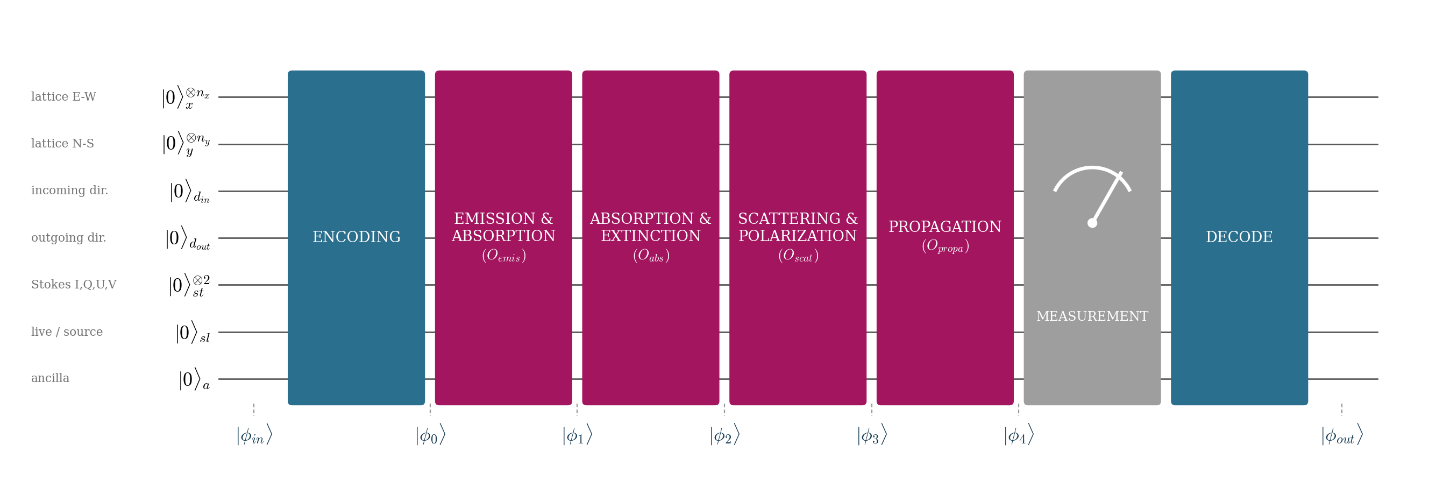}
\caption{Fig. 2. Schematics of quantum radiative transfer coding structure.}
\end{figure}

\subsubsection{Implementation of the Emission and Absorption Process}

The update of absorption radiative transfer process adds the emission source amplitudes encoded in the $\ {|1\rangle }_{s}$ subspace, to the stream amplitudes encoded in the$\ {|0\rangle }_{s}$ subspace. This process is described by the linear update in Eq.\,(6).

For $\phi_{1} = \begin{pmatrix}
S_{d}^{1} \\
J_{d}^{1}
\end{pmatrix}$ and $\phi_{0} = \begin{pmatrix}
S_{d}^{0} \\
J_{d}^{0}
\end{pmatrix}$:

\begin{equation}
\ \phi_{1} = O_{emis} \cdot \phi_{0} = O_{emis} \cdot \begin{pmatrix}
S_{d}^{0} \\
J_{d}^{0}
\end{pmatrix} = \ \begin{pmatrix}
1 & 1 \\
0 & 1
\end{pmatrix} \cdot \begin{pmatrix}
S_{d}^{0} \\
J_{d}^{0}
\end{pmatrix} = \begin{pmatrix}
S_{d}^{0} + J_{d}^{0} \\
J_{d}^{0}
\end{pmatrix}
\tag{6}
\end{equation}

Thus, $\text{\:}S_{d}^{1} = S_{d}^{0} + J_{d}^{0},$ $\text{\:}J_{d}^{1} = J_{d}^{0}$

which corresponds to putting the emission of the external source radiation (J) into the live stream (S) without depleting the source. Note that the subscript denotes the direction d, while the superscript denotes the time step t.

Since$\ O_{emis}$ is non-unitary, it cannot be implemented directly as a quantum gate. Instead, we realize $O_{emis}$ using a linear-combination-of-unitaries (LCU) block-encoding [16-18]. In particular, $O_{emis}$ admits the decomposition.

\begin{equation}
O_{emis} = I + \frac{1}{2}X + \frac{1}{2}ZX
\tag{7}
\end{equation}

where $I$, $X$ and $Z$ denote the identical, Pauli-$X$ and Pauli-$Z$ operators, which act on the stream/source qubit $ls$. The corresponding LCU implementation is realized using a two-qubit ancilla registers that control the application of the unitaries$\ \{ I,\ \ X,\ \ ZX\}$, followed by post-selection on the ancilla register. Accordingly, the post-selected update of emission and absorption processes and its equivalent expansion form is expressed by Equation 8, where $\left\| \phi \right\|$ is the $\mathcal{l}_{2}$-norm of $\phi$, while 2 is the norm of $O_{emis}$.

\begin{equation}
\begin{split}
\left|\phi_{1}\right\rangle 
={} & |00\rangle_{a}\,\frac{1}{2\|\phi_{0}\|}\sum_{i_{x},i_{y}=0}^{M-1}\sum_{d=0}^{n_{d}-1} \Big[\left(S_{d}^{(i_{x},i_{y})} + \Delta t\,J_{d}^{(i_{x},i_{y})}\right)|0\rangle_{sl} + \Delta t\,J_{d}^{(i_{x},i_{y})}|1\rangle_{sl}\Big] \\
    & \otimes |d_{in}\rangle_{in}|d_{out}\rangle_{out}|i_{x}\rangle_{x}|i_{y}\rangle_{y}
\end{split}
\tag{8}
\end{equation}

To visualize the construction of this gate, the corresponding two-ancilla circuit is illustrated in Fig. 3

\begin{figure}[htbp]
\centering
\includegraphics[width=3.76111in,height=1.24834in]{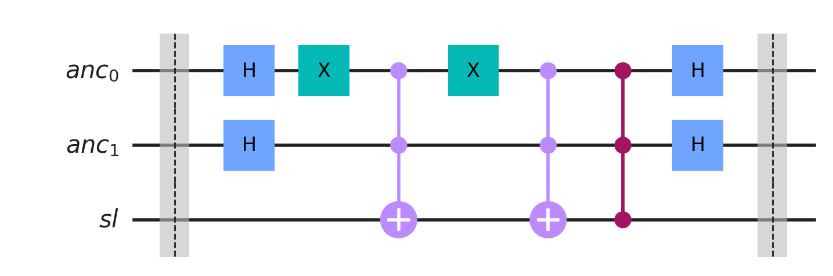}
\caption{Fig. 3: Visualization of the emission-absorption gate.}
\end{figure}

\subsubsection{Implementation of the Absorption and Extinction Processes}

From the transport equation (Eq. 2), the extinction coefficient accounts for both absorption and scattering, $k_{ext} = k_{abs} + k_{scat}$. Isolating the absorption contribution, the attenuation over one small time step is given by

\begin{equation*}
S_{d}^{n+1} = a_{abs}\,S_{d}^{n}, \qquad a_{abs} = e^{-k_{abs}(r)\Delta t} \approx 1 - k_{abs}(r)\Delta t
\end{equation*}

where $a_{abs}$ is absorption amplitude that remains in the live sector, or the Beer-Lambert transmission ($e^{- k_{abs}(r) \bigtriangleup t})$. In the study of Igarashi et al. (2024) the first order Taylor approximation, $1 - k_{abs}(r) \bigtriangleup t$ was used. $\sqrt{1 - {a_{abs}}^{2}}$ is amplitude transferred to loss (ancilla). The absorption update acts only on the live sector$|0\rangle_{sl}$, while the $|1\rangle_{sl}$ sector (source term) is left unchanged.

Because this mapping is also not unitary, it cannot be applied directly as a quantum gate. We embed it into a larger unitary operation using one ancilla qubit initialized in ${|0\rangle }_{a}$. And thus, the absorption block is implemented as a rotation $R_{y}$ (Eq. 9).

\begin{equation}
O_{abs} = \begin{pmatrix}
a_{abs} & -\sqrt{1 - a_{abs}^{2}} \\
\sqrt{1 - a_{abs}^{2}} & a_{abs}
\end{pmatrix} = R_{y}\!\left(2\arccos a_{abs}\right)
\tag{9}
\end{equation}

In this sense, the absorption dynamics are the complementary counterpart of the emission process: rather than adding amplitude into the live sector, it coherently transfers part of that amplitude into the ancilla branch associated with loss. The corresponding two-dimensional unitary acting on the ancilla subspace. Accordingly, the post-absorption extinction state is updated in Eq. 10.

\begin{equation}
\begin{split}
\left|\phi_{2}\right\rangle 
={} & \frac{1}{2\|\phi_{0}\|}\sum_{i_{x},i_{y}=0}^{M-1}\sum_{d=0}^{n_{d}-1}\left(\left(S_{d}^{(i_{x},i_{y})} + \Delta t\,J_{d}^{(i_{x},i_{y})}\right)|0\rangle_{sl} + \Delta t\,J_{d}^{(i_{x},i_{y})}|1\rangle_{sl}\right) \\
    & \times \Big[|0\rangle_{a}\otimes a_{abs}|0\rangle_{sl} + |1\rangle_{a}\otimes\sqrt{1-a_{abs}^{2}}\,|0\rangle_{sl}\Big] \otimes |d_{in}\rangle_{in}|d_{out}\rangle_{out}|i_{x}\rangle_{x}|i_{y}\rangle_{y}
\end{split}
\tag{10}
\end{equation}

And the absorption gate structure is shown in Figure 4.

\begin{figure}[htbp]
\centering
\includegraphics[width=3.95556in,height=1.69853in]{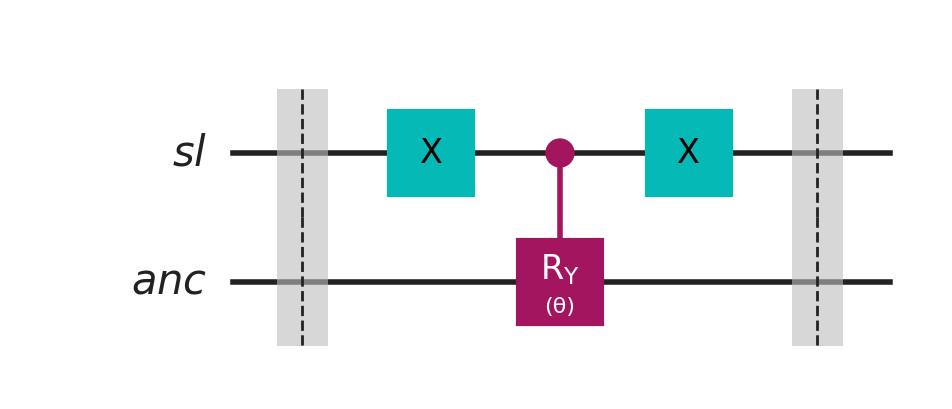}
\caption{Fig. 4. Visualization of the absorption-extinction gate.}
\end{figure}

\subsubsection{Implementation of the Scattering and Polarization Processes}

The scattering contribution consists of two physically distinct components:

(i) a survival (no-scatter) term, which leaves the photon in its current direction, and

(ii) a redistribution and depolarization term, which transfers amplitude into other directions according to the scattering phase function.

From the discrete transport equation, the scattering operator can be written as the sum of scattering-out and scattering-in contributions,

\begin{equation}
S_{d}^{n+1} = \underbrace{(1 - k_{scat}\Delta t)\,S_{d}^{n}}_{\text{surv}} + \underbrace{k_{scat}\Delta t\sum_{d^{'}=0}^{n_{d}-1}\omega_{d^{'}}P(d,d^{'})\,S_{d^{'}}^{n}}_{\text{redist}}
\tag{11}
\end{equation}

${(1 - k}_{scat}\Delta t{)S}_{d}^{n}$ represents photons that remain unscattered in direction $d$, while the second term ($k_{scat}\Delta t\sum_{d^{'} = 0}^{Nd}{\omega_{d^{'}}P\left( d,d^{'} \right)S_{d^{'}}^{n}})\ $redistributes intensity from all incoming directions $d^{'}$into direction $d$ according to phase function, specifically the Rayleigh phase function for gas scattering.

Although $O_{scat}$is not unitary, unitarity can be recovered by embedding this operator into a larger unitary evolution using the linear-combination-of-unitary (LCU) framework [17].

So, the decomposition in physics is

\begin{equation}
O_{scat} = \underset{\alpha_{surv}}{\overset{(1 - k_{scat}\Delta t)}{}}I + \underset{\alpha_{redist}}{\overset{k_{scat}\Delta t}{}}\ R_{Rayleigh}
\tag{12}
\end{equation}

where $I$ is the identity and $R_{Rayleigh}$ is the redistribution kernel.

\paragraph{Redistribution of Scattering Light -- Single In-Out Scattering Angle Pair}

For a fixed pair (dout, din), the polarization transformation relating the incident and scattered Stokes vectors is constructed as a rotation--scattering--rotation sequence, that is, the rotation of scattering plane from scattering surface to meridian surface, and the Rayleigh phase function and meridian surface to scattering surface (see, Eq. 13). Throughout this process, the Stokes parameters ($I,\ Q,U\ ,V)$ are referred to the local meridian plane that contains the propagation direction and the polar (stratification) axis $\widehat{\mathbf{z}}$ whereas the Rayleigh scattering matrix is naturally defined in the scattering plane, spanned by $d_{in}$and $d_{out}$. As classic light scattering consideration, since these reference frames do not coincide, the incident Stokes vector is first rotated (R) from the incoming meridian plane into the scattering plane, the phase matrix (P) of Rayleigh is applied, and the resulting vector is rotated (R) from the scattering plane back into the outgoing meridian plane, namely, Rotation--Phase(scattering)--Rotation (RPR) sequence (Eq. 13). The updated Stokes vector is therefore written as:

\begin{equation}
S_{d}^{n + 1}\  = R\left( \psi_{out} \right)P(\theta)R\left( \psi_{in} \right)S_{d}^{n}
\tag{13}
\end{equation}

where $\theta$  is the scattering angle, and $\psi_{in}$and $\psi_{out}$ are the input and output frame-rotation angles derived from the incident and outgoing directions. Geometrically,

\begin{equation}
\theta = arccos({\widetilde{v}}_{out} \bullet {\widetilde{v}}_{in})
\tag{14}
\end{equation}

and the polarization-frame angles are computed from the corresponding directional geometry:

$\psi_{in} = atan2({\widetilde{v}}_{in,y},\ {\widetilde{v}}_{in,\ x})$, 
\begin{equation}
\psi_{out} = atan2({\widetilde{v}}_{out,y},\ {\widetilde{v}}_{out,\ x})
\tag{15}
\end{equation}

Thus, same as classical way, the quantum polarization update is also implemented as a direction-dependent Stokes-space RPR sequence that includes both scattering phase function as well as scattering plane rotation using Eq 13.

For single scattering phase function of Rayleigh $PF(\theta)$ at scattering angle $\theta$ in Eq. 13, the transformation of polarization is governed by the Mueller matrix $P(\theta)$ (Eq. 16) and preserves conservation (Eq. 17)

\begin{equation}
P(\theta) = \frac{3}{8\pi}\begin{pmatrix}
\begin{matrix}
\frac{1 + \cos^{2}\theta}{2} & - \frac{\sin^{2}\theta}{2} \\
 - \frac{\sin^{2}\theta}{2} & \frac{1 + \cos^{2}\theta}{2}
\end{matrix} & \begin{matrix}
0 & 0 \\
0 & 0
\end{matrix} \\
\begin{matrix}
0 & 0 \\
0 & 0
\end{matrix} & \begin{matrix}
cos\theta & 0 \\
0 & cos\theta
\end{matrix}
\end{pmatrix} = \frac{3}{8\pi}\widetilde{P}(\theta)
\tag{16}
\end{equation}

\begin{equation}
\omega_{d_{out},{\ d}_{in}} = \ \frac{P\left( \theta_{d_{out},{\ d}_{in}} \right)\mathrm{\Delta}\omega_{d_{out}}}{\sum_{d_{out} = k}^{}{P\left( \theta_{d_{out},{\ d}_{in}} \right)\mathrm{\Delta}\omega_{d_{out}}}},\ \ \sum_{d_{out} = k}^{}\omega_{d_{out},{\ d}_{in}} = 1
\tag{17}
\end{equation}

here, the Mueller matrix $P(\theta)$ is the Rayleigh phase function, and $\widetilde{P}(\theta)$ is normalized phase function, where $\parallel \widetilde{P}(\theta) \parallel \leq 1$. $\widetilde{P}(\theta)$ is real and symmetric but not unitary, so it cannot be realized directly as a quantum gate.

We therefore embed it as the top-left block of a larger unitary
\begin{equation*}
U_{\widetilde{P}(\theta)} = \begin{pmatrix}
\widetilde{P}(\theta) & \sqrt{I - {\widetilde{P}(\theta)}^{2}} \\
\sqrt{I - {\widetilde{P}(\theta)}^{2}} & - \widetilde{P}(\theta)
\end{pmatrix}
\end{equation*}
(which is a $8 \times 8$ matrix), where $\widetilde{P}(\theta)$ can be obtained by acting on one ancilla qubit (Eq. 18).

\begin{equation}
\widetilde{P}(\theta) = \ \langle 0_{a}|U_{\widetilde{P}(\theta)}|0_{a}\rangle
\tag{18}
\end{equation}

The relationship of $\widetilde{P}(\theta)$ and its eigenvalue satisfies $\det{\left( \widetilde{P}(\theta) - \lambda_{k}I \right)v} = 0$ where v is a nonzero vector. Splitting the normalized operator
\begin{equation*}
\widetilde{P}(\theta) = \begin{pmatrix}
a & b & 0 & 0 \\
b & a & 0 & 0 \\
0 & 0 & c & 0 \\
0 & 0 & 0 & c
\end{pmatrix}
\end{equation*}
(where $a = a = \ (1 + \cos^{2}\theta)/2$, $b = \ {- \sin}^{2}\theta/2$, c=$\cos\theta$ ) into two pieces (the I/Q part, top-left $2 \times 2$ matrix and the U/V part, bottom-right $2 \times 2$ matrix), for a block-diagonal matrix, the determinant is the product of the blocks' determinants
\begin{equation*}
\det\left( \widetilde{P}(\theta) - \lambda_{k}I \right) = \det\!\begin{pmatrix}
a - \lambda_{k} & b \\
b & a - \lambda_{k}
\end{pmatrix} \times \det\!\begin{pmatrix}
c - \lambda_{k} & 0 \\
0 & c - \lambda_{k}
\end{pmatrix} = 0.
\end{equation*}
and hence the eigenvalues $\lambda_{k} = \left\{ 1,{\cos}^{2}\theta,\cos\theta,\cos\theta \right\}$. Each eigenvalue ($\lambda_{k}$) is mapped to a single-qubit rotation using the $R_{y}$ gate (UCRYGate)
\begin{equation*}
R_{Y}(\theta_{k}) = \begin{pmatrix}
\cos\frac{\theta_{k}}{2} & -\sin\frac{\theta_{k}}{2} \\
\sin\frac{\theta_{k}}{2} & \cos\frac{\theta_{k}}{2}
\end{pmatrix} = \begin{pmatrix}
\lambda_{k} & -\sqrt{1 - \lambda_{k}^{2}} \\
\sqrt{1 - \lambda_{k}^{2}} & \lambda_{k}
\end{pmatrix}
\end{equation*}
acting on the ancilla qubit ($\langle 0|R_{Y}\left( \theta_{k} \right)|0\rangle = \lambda_{k}$), with rotation angles determined by

\begin{equation}
\theta_{k} = \ 2arccos(\lambda_{k})
\tag{19}
\end{equation}

Concretely, the circuit applies no rotation for the invariant mode ($\lambda_{1} = 1$), a rotation $2arccos({cos}^{2}\theta)$ to the parallel mode ($\lambda = {cos}^{2}\theta$), and equal rotations $2arccos(cos\theta)$ to the $U$ and $V$ components.

To reconstruct the Stokes vector, after the controlled rotations, the basis transformation is uncomputed (Eq. 20) to return to the original Stokes representation.

$U_{\widetilde{P}(\theta)} = {V^{\dagger}UCR}_{y}(\theta_{1},\ \theta_{2},\ \theta_{3},\ \theta_{4})V$ 
\begin{equation}
\ \ 
\tag{20}
\end{equation}

here, $U_{\widetilde{P}(\theta)}$ acts on the Stokes registers together with the ancilla. V denotes the basis transformation that rotates the Stokes-space operator into the eigen basis of $\widetilde{P}(\theta)$, which is the Hadamard basis here (applied as H to st[0], namely I and Q), and its adjoint $V^{\dagger}$ rotates back to the Stokes representation once the controlled rotations have rotated the ancilla by the angle corresponding to each Stokes component.

Because the Stokes parameters $Q$ and $U\ $depend on the choice of polarization reference axes, any scattering update must account for rotations of the polarization frame. A rotation of the polarization basis by an angle $\psi$ is described by the Mueller rotation matrix

R($\psi$) = 
\begin{equation}
\begin{pmatrix}
\begin{matrix}
1 & 0 \\
0 & \cos 2\psi
\end{matrix} & \begin{matrix}
0 & 0 \\
\sin 2\psi & 0
\end{matrix} \\
\begin{matrix}
0 & - \sin 2\psi \\
0 & 0
\end{matrix} & \begin{matrix}
\cos 2\psi & 0 \\
0 & 1
\end{matrix}
\end{pmatrix}
\tag{21}
\end{equation}

Under this transformation, the Stokes vector (I, Q, U, V) becomes

\begin{equation}
\begin{pmatrix}
\begin{matrix}
I \\
Q
\end{matrix} \\
\begin{matrix}
U \\
V
\end{matrix}
\end{pmatrix}\overset{R(\psi)}{\rightarrow}\begin{pmatrix}
\begin{matrix}
I \\
Qcos2\psi + Usin2\psi
\end{matrix} \\
\begin{matrix}
Ucos2\psi - Qsin2\psi \\
V
\end{matrix}
\end{pmatrix}
\tag{22}
\end{equation}

The appearance of the doubled angle $2\psi$ reflects the second-rank tensor character of linear polarization.

In the circuit implementation, this rotation is realized by the two-qubit operator generated by

\begin{equation}
O_{R} \equiv X \otimes X + Y \otimes Y = \ R_{XX + YY}
\tag{23}
\end{equation}

which coherently mixes the $\left| 01 \right\rangle$ (Q) and $|10\rangle$ (U) subspace while leaving $|00\rangle$ (I) and $|11\rangle$ (V) unchanged. which in Qiskit is realized using the XXPlusYYGate as

\begin{equation}
R_{XX + YY}(\theta,\ \beta) = \ \ \begin{pmatrix}
\begin{matrix}
1 & 0 \\
0 & \cos\frac{\theta}{2}
\end{matrix} & \begin{matrix}
0 & 0 \\
 - isin\frac{\theta}{2}e^{- i\beta} & 0
\end{matrix} \\
\begin{matrix}
0 & - isin\frac{\theta}{2}e^{i\beta} \\
0 & 0
\end{matrix} & \begin{matrix}
\cos\frac{\theta}{2} & 0 \\
0 & 1
\end{matrix}
\end{pmatrix}\ ;
\tag{24}
\end{equation}

To get Equation 21, hence $\theta = 4\psi;\ \ \beta = - \pi/2$ needs to be signed to the XXPlusYYGate.

The design of the single in--out scattering RPR gate is shown in Figure 5.

\begin{figure}[htbp]
\centering
\includegraphics[width=5.29444in,height=1.99107in]{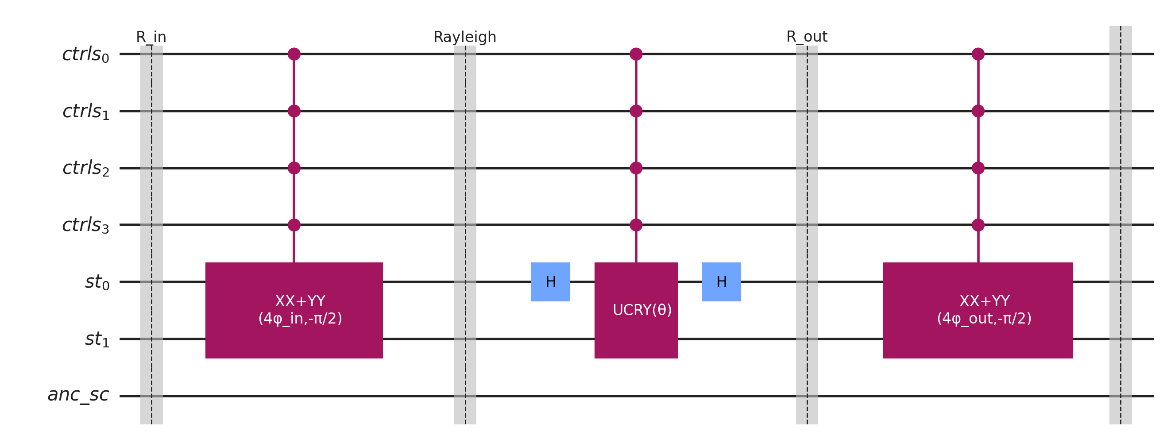}
\caption{Fig. 5. Visualization of single directional RPR gate.}
\end{figure}

\paragraph{Sum of All Scattering Direction Pairs}

The RPR gate described above realizes the scattering contribution of a single ordered direction pair ($d_{out}$, ${\ d}_{in}$). To obtain the full outgoing radiation field, however, this operation must be applied coherently to every pair and the contributions summed over all incoming directions, yielding the directionally-resolved Stokes vector for each outgoing direction (Eq. 25).

\begin{equation}
A = \sum_{d_{out} = 0}^{n_{d} - 1}{\omega_{d_{out},{\ d}_{in}}U_{d_{out},{\ d}_{in}}}
\tag{25}
\end{equation}

A is weighted sum of the per-pair RPR unitaries, with non-negative coefficients $\omega_{d_{out},{\ d}_{in}}$. Although each pair is unitary, the sum is not and could not be applied directly in quantum circuit. Conceptually, this sum must be realized following the standard Linear Combination of Unitary (LCU) framework [17]. The LCU trick embeds it as a block of a larger unitary using one extra "routing" register and be presented as the PREP--SELECT--PREP$\dagger$  sandwich [16-17]. This sandwich construction is described below in 3 stages sequence.$\text{PREP}_{\text{joint}}\text{\:\,} \rightarrow \text{\:\,}\text{SELECT}_{\text{RPR}}\text{\:\,} \rightarrow \text{\:\,}\text{PREP}_{\text{joint}}^{\dagger}$. $\text{PREP}_{\text{joint}}$ is the joint direction preparation (see stage 1 below), which encodes the weight ($\omega_{d_{out},{\ d}_{in}}$, Eq. 25) onto the route register, a particular scattering route; $\text{SELECT}_{\text{RPR}}$ is the Multiplexed selection (see stage 2 below), which applies the directional RPR operator $U_{d_{out},{\ d}_{in}}$ in each LCU branch through controlled operations; and $\text{PREP}_{\text{joint}}^{\dagger}$ is inverse of $\text{PREP}_{\text{joint}}$ that uncomputes the direction selection, recombines the branches through interference and restores the routing register to $|0\rangle$ (see Stage 3 below). This construction ensures that the appropriate single scattering rotation of directional Rotation--Phase(scattering)--Rotation (RPR, Eq. 13) operator is applied conditionally to each ordered pair of incoming and outgoing directions, while preserving overall unitarity.

Stage 1: Joint Direction Preparation ($\text{PREP}_{\text{joint}}$) -- Encode the Weights

When radiation from in an incoming direction $d_{\text{in}}$ scatters, it is redistributed over all outgoing directions according to the scattering phase function (Eq. 16), and this redistribution conserves energy (Eq. 17). Consequently, the scattering of a single incoming beam is a weighted sum of the single-pair RPR operators (Eq. 25). However, the sum of unitaries is not itself unitary and can\textquotesingle t be applied directly, so it is realized through the linear-combination-of-unitaries (LCU) construction. The first step of that construction, the joint preparation operator $\text{PREP}_{\text{joint}}$, loads these weights (Eq. 25) onto the routing register (the ancilla register that indexes the outgoing directions) by writing the square-root coefficients (normalization) as the amplitudes of a coherent superposition. These amplitudes are the LCU weights that the SELECT step (Stage 2) will use.

Acting on the initial ancilla state $\mid 0\rangle_{a}$, PREP produces (Eq. 26)

$\text{PREP}_{\text{joint}}|0\rangle_{a}\text{\:\,} \longrightarrow \text{\:\,}\sum_{d_{out} = 0}^{n_{d} - 1}{\sqrt{\frac{\omega_{d_{out},{\ d}_{in}}}{\lambda}}\text{\,}|d_{out}\rangle_{a}}$, 
\begin{equation}
\lambda = \sum_{d_{out} = 0}^{n_{d} - 1}{\omega_{d_{out},{\ d}_{in}} = 1}
\tag{26}
\end{equation}

where each basis state $\mid d_{out}\rangle_{a}$ labels one outgoing-direction branch, the weights $\omega_{d_{out},{\ d}_{in}} > 00$~are the LCU coefficients, and $\lambda$ is the normalization factor equal to their sum ($\lambda = 1$ for the column-normalized phase-function weights). Writing the amplitudes as square-root guarantees that the prepared state is properly normalized, since the squared amplitudes $\frac{\omega_{d_{out},{\ d}_{in}}}{\lambda}$ sum to one.

Conceptually, $\text{PREP}_{\text{joint}}$ places the ancilla in a coherent superposition of all $n$ outgoing directions, with each branch carrying amplitude equal to the square root of its scattering weight. Because the superposition is coherent rather than a classical choice of a single direction, the controlled operations of the next stage act on every directional branch at once, so all outgoing directions are "live" simultaneously. The inverse preparation in Stage 3 then recombines these branches through interference: the branch that returns the ancilla to $|0\rangle_{a}$collects exactly the weighted sum $\sum_{d_{out}}^{}\omega_{d_{out},{\ d}_{in}}U_{d_{out}}$. This interference is what assembles the superposition into the desired LCU, allowing all directional scattering contributions to be added coherently in a single pass.

The routing register requires $\left\lceil {log}_{2}n \right\rceil$ qubits (ancilla qubits) to index the $n\ $directions, and the weight encoding within ($\text{PREP}_{\text{joint}}\ $ is realized as a structured binary tree of $R_{Y}$ rotations. The construction proceeds recursively over the routing qubits in most-significant-first order. At each level the current qubit receives an $R_{Y}$ rotation, conditioned on the qubits already fixed above it, that divides the remaining probability mass between its $0$-branch and $1$-branch in proportion to the summed weights of the two halves; the routine then descends into each half with the relative (renormalized) weights, conditioned on that qubit\textquotesingle s value. The rotation angle at each node is $\theta = 2arccos\sqrt{p_{\text{left}}/p_{\text{node}}}$, chosen so that the squared amplitudes reproduce the target proportions. Because the per-level square-root factors multiply telescopically along each root-to-leaf path, the final amplitude on direction $d_{out}$ is exactly $\sqrt{w_{d_{out}\text{\,}d_{\text{in}}}}$. Only $R_{Y}$ rotations are required, since the scattering weights are real and non-negative, and the resulting tree has depth $\left\lceil {log}_{2}n \right\rceil$ for $n$ directions.

Stage 2: Multiplexed selection ($\text{SELECT}_{\text{RPR}}$) --Apply the Directional RPR Operators

A multiplexed selection operator, denoted by $\text{SELECT}_{\text{RPR}}$, conditionally applies one of~n~RPR unitaries according to the state of the ancilla register. Within the LCU framework, it applies every directional operator to its corresponding branch of the superposition simultaneously. For an n-term LCU expansion, the selection operator is defined as

\begin{equation}
\text{SELECT}_{\text{RPR}} = \ \sum_{d_{out} = 0}^{n_{d} - 1}{|d_{out}\rangle{\langle d_{out}|}_{a} \otimes U_{d_{out},{\ d}_{in}}}
\tag{27}
\end{equation}

When acting on the prepared ancilla superposition generated by~$\text{PREP}_{\text{joint}}$ the combined transformation becomes:

\begin{equation}
\text{SELECT}_{\text{RPR}}\left( \text{PREP}_{\text{joint}}|0\rangle_{a}|\phi_{2}\rangle \right)\text{\:\,} = \text{\:\,}\sum_{d_{out},{\ d}_{in} = 0}^{n_{d} - 1}{\sqrt{\frac{\omega_{d_{out},{\ d}_{in}}}{\lambda}}|d_{out}\rangle_{a} \otimes U_{d_{out},{\ d}_{in}}|\phi_{2}\rangle}
\tag{28}
\end{equation}

Here, each unitary ~$U_{d_{out},{\ d}_{in}}\ $ is the RPR operator associated with a specific incoming--outgoing direction pair. Together with the Stage 1 preparation, in which the routing register coherently encodes all directional branches with amplitudes equal to the square roots of the normalized LCU coefficients, SELECT applies each RPR operator to its corresponding branch. Consequently, the system evolves into a coherent superposition of all directional RPR transformations, enabling quantum interference among the different propagation pathways.

In the $\text{SELECT}_{\text{RPR}}$ stage of Figure 6, the outgoing-direction register ${|d}_{out}\rangle$ is computed in place from the routing register before the directional operator is applied. Both blocks ($\text{PREP}_{\text{joint}}$, $\text{SELECT}_{\text{RPR}}$) act by bitwise XOR (addition modulo 2 on the direction-qubit strings) and fire only on the scattering branch (${anc}_{\text{sel}} = 1$, $\text{sl} = 0$)) On entry, the outgoing register holds a copy of the incoming direction, $d_{out} = d_{in}$. The first block ($d_{out}{\bigoplus = d}_{in}$), therefore reset it to zero ( $d_{out} = d_{in} \oplus d_{in} = 0$). The second block ($d_{out}\bigoplus$=route) then writes the routing selection into it ($d_{out} = \ 0 \oplus k = k$), where k is the outgoing direction prepared on the route register by the $\text{PREP}_{\text{joint}}$ stage. After these two operations the outgoing register labels the scattered direction of each branch, $d_{out} = k$, while the route register still holds $k$and is uncomputed afterward by $\text{PREP}_{\text{joint}}^{\dagger}$ (valid because $d_{in}$ is unchanged through $\text{SELECT}_{\text{RPR}}$. Conditioned on the resulting incoming--outgoing pair ($d_{out}$, $d_{in}$), the directional RPR operator U($d_{out}$, $d_{in}$) = $R_{out}\widetilde{P}R_{in}$ is applied to the Stokes register: rotates the Stokes vector into the scattering plane of the incoming direction, $\widetilde{P}$ applies the Rayleigh phase matrix, and $R_{out}$ rotates it into the outgoing frame. Because XOR is its own inverse, the in-place direction update is fully reversible.

Stage 3: Uncomputation ($\text{\,}\text{PREP}_{\text{joint}}^{\dagger}$) -- Recombine the Branches

Finally, the joint preparation is uncomputed to disentangle and reset the routing workspace. This process is the inverse of the preparation step $\text{PREP}_{\text{joint}}^{\dagger}$, which coherently recombines the directional branches and~rotates the ancilla register back toward the reference state $|0\rangle_{a}$. The inverse runs though the RY-tree backwards: every $R_{Y}(\theta)$ becomes $R_{Y}( - \theta)$ and the gate order is reversed. Through quantum interference, the weighted unitary contributions add coherently to realize the desired linear combination. Applying $\text{\,}\text{PREP}_{\text{joint}}^{\dagger}$, the entangled ancilla--system state yields:

\begin{equation}
\text{PREP}_{\text{joint}}^{\dagger}\text{SELECT}_{\text{RPR}}\text{PREP}_{\text{joint}}|0\rangle_{a}|\phi_{2}\rangle = |0\rangle_{a} \otimes \frac{1}{\lambda}\sum_{d_{out} = 0}^{n_{d} - 1}\omega_{d_{out},{\ d}_{in}}U_{d_{out},{\ d}_{in}}|\phi_{2}\rangle + \mid \bot\rangle_{a}
\tag{29}
\end{equation}

Here, $\mid \bot\rangle_{a}$denotes ancilla states orthogonal to $|0\rangle_{a}$, These orthogonal components contain residual ``junk'' information arising from imperfect overlap between the interfering branches and do not contribute to the successful LCU outcome. Since conditioning on the ancilla are returned to $|0\rangle_{a}$ implementations therefore reach the desired operator

\begin{equation}
\left( \text{\,} \middle| 0\rangle_{a} \otimes I \right)\text{ PREP}_{\text{joint}}^{\dagger}\text{SELECT}_{\text{RPR}}\text{PREP}_{\text{joint}}\ \left( |0\rangle_{a} \otimes I \right) = \ \frac{1}{\lambda}\sum_{d_{out} = 0}^{n_{d} - 1}{\omega_{d_{out},{\ d}_{in}}U_{d_{out},{\ d}_{in}}}\text{\:}
\tag{30}
\end{equation}

Thus, the overall $\text{PREP}_{\text{joint}}$--$\text{SELECT}_{\text{RPR}}$--$\text{PREP}_{\text{joint}}^{\dagger}$~sequence (Figure 6) realizes the weighted linear combination of directional RPR unitaries within a coherent quantum circuit. The ancilla preparation distributes amplitudes across the directional branches, the $\text{SELECT}_{\text{RPR}}$stage applies the corresponding controlled unitaries, and the final uncomputation step interferometrically recombines the branches to produce the target operator.

\begin{figure}[htbp]
\centering
\includegraphics[width=5.29097in,height=2.05659in]{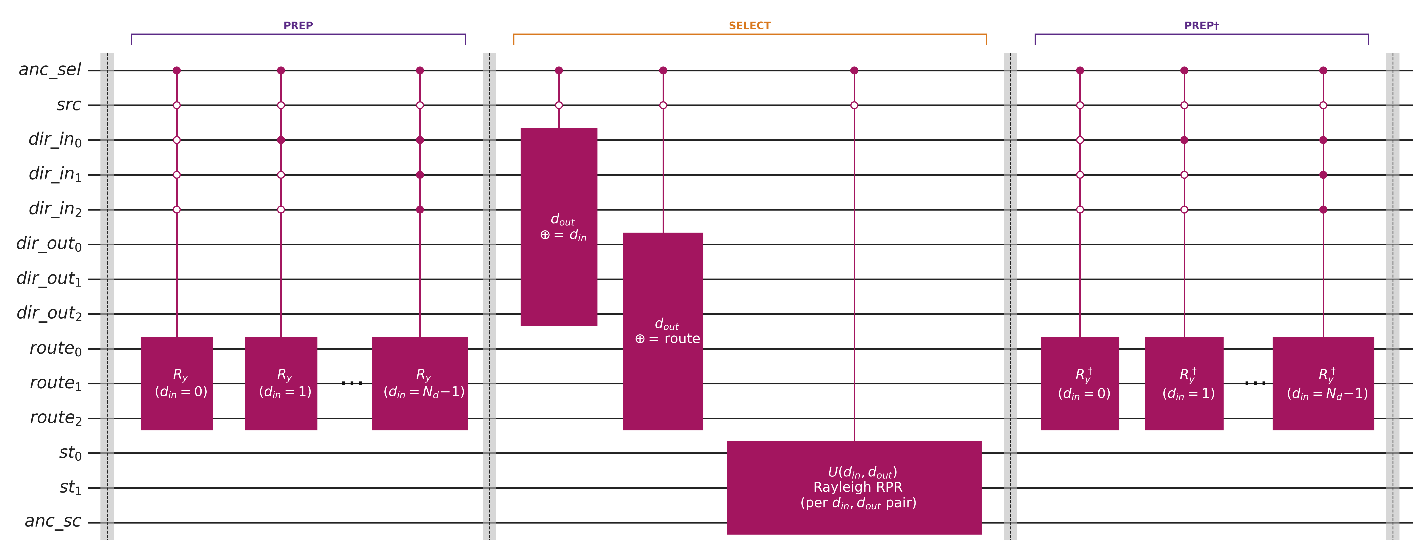}
\caption{Fig. 6: Visualization of the $\text{PREP}_{\text{joint}}\text{\:\,} \rightarrow \text{\:\,}\text{SELECT}_{\text{RPR}}\text{\:\,} \rightarrow \text{\:\,}\text{PREP}_{\text{joint}}^{\dagger}$ integration gate}
\end{figure}

\paragraph{LCU Integrations of Survival and Redistributed Scattering Components}

In the previous sections we showed how to build the gate for the redistributed (scattered) component. However, the scattering of live radiation has two parts: a survival component, in which the radiation continues unchanged, and a redistributed component, in which it is scattered into new directions. The full scatter-on-live operator is therefore the weighted sum $A = \alpha_{0}I + \ \alpha_{1}S_{redist}$ Because this sum of unitaries is non-unitary, an ancilla-assisted construction is employed. The selector ancilla, which is the branch ancilla of this two-term LCU and selects between the survival and redistribution branches, is prepared using a single-qubit unitary of the form

\begin{equation}
\begin{split}
U_{prep}\left|0\right\rangle 
={} & \sqrt{\frac{\alpha_{surv}}{\lambda}}\left|0\right\rangle + \sqrt{\frac{\alpha_{redist}}{\lambda}}\left|1\right\rangle \\
={} & R_{y}\!\left(2\arctan\sqrt{\frac{\alpha_{redist}}{\alpha_{surv}}}\right)
\end{split}
\tag{31}
\end{equation}

where $\alpha_{surv}$is the survival weight, $\alpha_{scat}$is the scattering weight, and $\lambda = \alpha_{surv} + \alpha_{scat}$. The branch $|0\rangle $ corresponds to the survival branch, while the branch $|1\rangle $ corresponds to the redistributed scattering branch. This LCU gate is shown in Figure 7.

\begin{figure}[htbp]
\centering
\includegraphics[width=4.85556in,height=3.15112in]{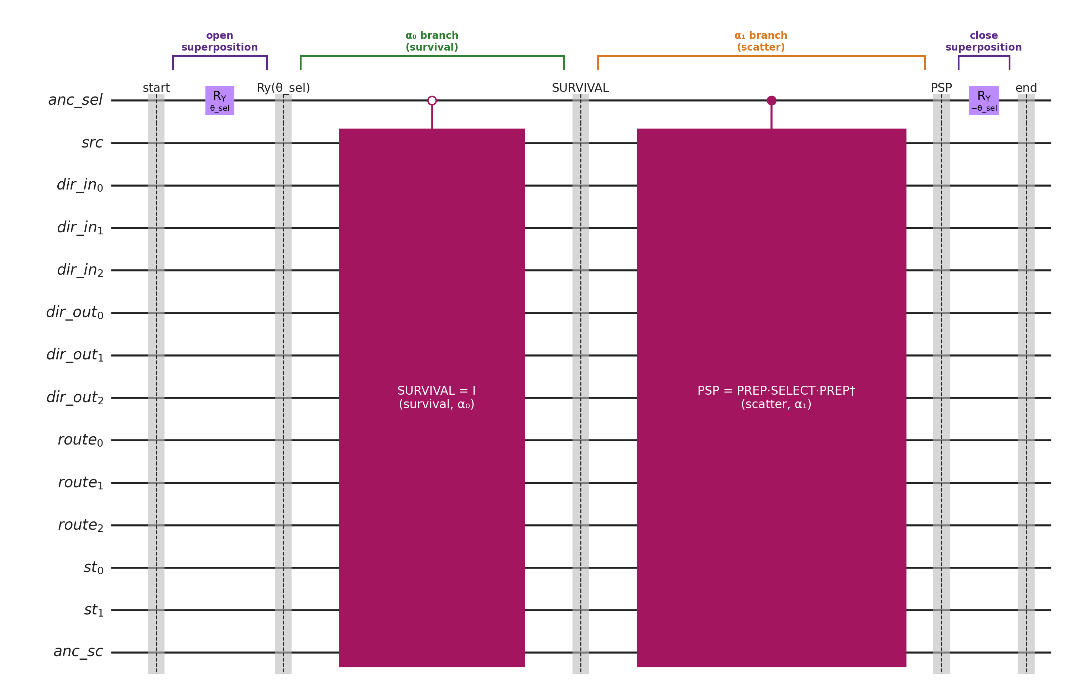}
\caption{Fig. 7: LCU of the survival and redistributed scattering gate}
\end{figure}

In summary, within the full circuit the scattering operator is controlled by the source qubit and the direction register, so that the polarization scattering update is applied only to the live radiative sector and only for the occupied direction state. The operator is structured as a selector-based sum of a survival branch and a redistributed-scatter branch. The scatter branch is itself realized by an ancilla-assisted block-encoding, while the survival branch acts as the identity on the current diagonal live direction. For redistributed scatter branch, the RPR components are accumulated over all directions through the directional LCU construction ($\text{PREP}_{\text{joint}}\text{\:\,} \rightarrow \text{\:\,}\text{SELECT}_{\text{RPR}}\text{\:\,} \rightarrow \text{\:\,}\text{PREP}_{\text{joint}}^{\dagger}$).

\paragraph{Quantum Post-Selection}

After uncomputation of the selector and scatter workspaces, an effective live update appears in the all-zero ancilla sector.

\begin{equation}
\begin{split}
\left|\phi_{3}\right\rangle 
={} & |0\dots0\rangle_{a}\,\frac{1}{\|\phi_{0}\|}\sum_{i_{x},i_{y}=0}^{M-1}\sum_{d_{in},d_{out}}^{n_{d}} S_{d_{in},d_{out}}^{(i_{x},i_{y})}\bigg(|0\rangle_{a}\otimes R_{out}(d_{out})\widetilde{P}(\theta_{d_{in},d_{out}})R_{in}(d_{in}) \\
    & \qquad + |1\rangle_{a}\otimes R_{out}(d_{out})\sqrt{1-\widetilde{P}(\theta_{d_{in},d_{out}})\widetilde{P}(\theta_{d_{in},d_{out}})^{\dagger}}R_{in}(d_{in})\bigg) \\
    & \otimes |0\rangle_{sl}|d_{out}\rangle_{out}|d_{in}\rangle_{in}|i_{x}\rangle_{x}|i_{y}\rangle_{y}
\end{split}
\tag{32}
\end{equation}

\subsubsection{Implementation of the Propagation (Streaming) Process}

Having discussed how to handle the emission-absorption, extinction-absorption, and extinction-scattering processes within a single lattice site, we now need to implement how the radiation propagates to neighboring sites through propagation operator. The propagation (can be also called streaming, or adventure, Figure 8) operator is a reversible, direction-controlled lattice translation. Let l donate the lattice index register (an $n$-qubit register indexing $M = 2^{n}$ sites). Propagation is enabled only when \emph{sl} = 0 (live radiation); for \emph{sl} = 1, the operator acts as the identity, so the freshly injected source is not directly transported within the same step.

In one dimension, the operator performs a modular shift of the lattice register conditioned on source qubit $ss$ and the propagation direction $d$. For sl = 0 and d = 1 (right side of barrier, rightward propagation), a controlled increment $l \mapsto l + 1$ (mod M) is applied using a ripple-carry cascade of multi-controlled-NOT gates [18]. For $sl = 0$ and $d = 0$ (leftward propagation), a controlled decrement $l \mapsto l - 1$ (mod M) is applied via a reversible (borrow-propagation) circuit. All auxiliary operations are uncomputed, ensuring the operator remains unitarity.

\begin{figure}[htbp]
\centering
\includegraphics[width=5.43614in,height=4.16667in]{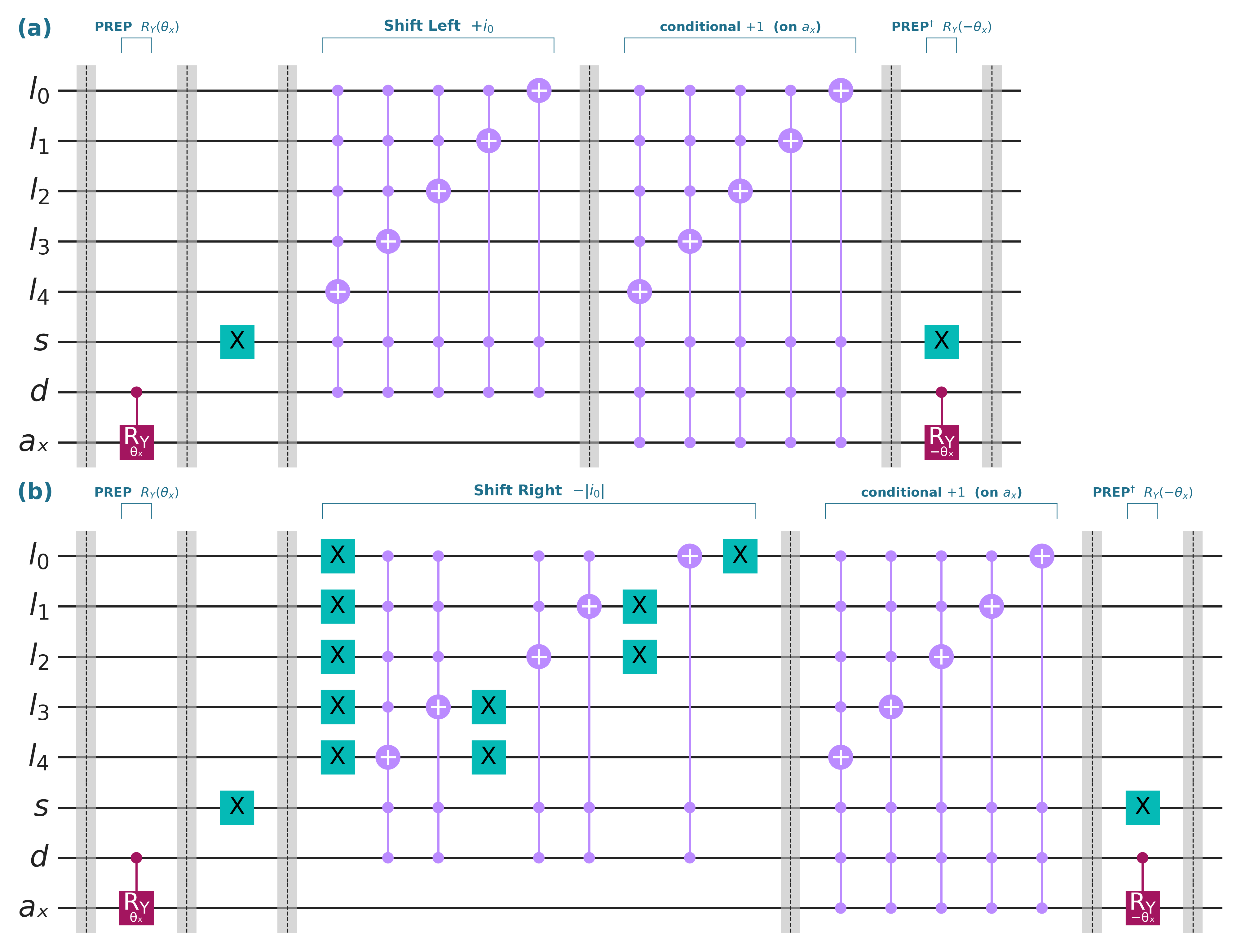}
\caption{Fig. 8: Visualization of the weighted propagation or streaming gate with lattice grids.}
\end{figure}

In two spatial dimensions, the azimuth of scattering direction determines how it propagates on the D2Q9 lattice, requiring a mapping from directional to advection information. The direction register $d$ encodes a discrete velocity $\mathbf{v}(d) = (v_{x}(d),v_{y}(d))$. Over one-time step the displacement in lattice units is

$\mathbf{l}(d) = (\frac{\ \mathbf{v}(d)\Delta t}{\Delta x},\ \ \frac{\ \mathbf{v}(d)\Delta t}{\Delta y}) = \ (l_{x},l_{y}),$ (33)

which is in general not an integer number of cells. Existing quantum LBM streaming operators are integer permutations of the lattice register [10, 19]. Here we lift this restriction by realizing the classical bilinear semi-Lagrangian scheme [20] as a linear combination of integer shifts on the quantum register. Since a lattice translation $T:\ f_{x}\  \mapsto \ f_{x - c}$ is defined only for integer ($n_{x} = int\left( l_{x} \right),n_{y} = int(l_{y})$), the streaming operator is constructed as a weighted sum of such translations to represent a real 2D adventure,

\begin{equation}
\mathbf{L}(d) = \sum_{b_{x},b_{y} \in \{ 0,1\}}^{}{\omega_{x}(b_{x}){\omega_{y}(b}_{y})T_{(i_{0} + b_{x},j_{0} + b_{y})}}
\tag{34}
\end{equation}

where $\omega_{x}(0) = 1 - \ f_{x}$, $\omega_{x}(1) = f_{x}\ $, $\omega_{y}(0) = 1 - f_{y}\ ,\ \omega_{y}(1) = f_{y}$, and the fractional part $f_{x} = l_{x} - n_{x}\ ,f_{y} = l_{y} - n_{y}$, and $0 < f_{x},\ f_{x} < 1$

Because the weights factories over the two axes, each axis is handled independently with a single ancilla qubit. For the x axis, the ancilla is prepared by an $R_{Y}$ rotation,

\begin{equation}
|\langle{0|}_{a_{x}}|\phi\rangle|^{2}\  = \ 1\  - \ f_{x},\ \ \ \ \ |\langle{1|}_{a_{x}}|\phi\rangle|^{2}\  = \ f_{x}
\tag{35}
\end{equation}

The full streaming unitary for one axis is then

$U_{propa}^{x}\  = \ R_{Y}^{a_{x}}( - \theta_{x})\ ({\left| 0 \right\rangle\left\langle 0 \right|}_{a_{x}} \otimes \ I_{a_{x}} + \ {\left| 1 \right\rangle\left\langle 1 \right|}_{a_{x}} \otimes \ T_{1}^{(x)})(\ I_{a_{x}} \otimes T_{i_{0}}\ )R_{Y}^{a_{x}}(\theta_{x})$ (36)

Where $\theta_{x}(d) = 2\ arcsin\sqrt{f_{x}(d)}$, $T_{i_{0}}$ is the base translation by $i_{0}$ = $n_{x}$ cells, implemented as a controlled increment or decrement depending on the sign, and $T_{1}$ conditioned on $|1\rangle_{a}$ is a unit translation applied when the ancilla is in state $|1\rangle$. On the postselected $|0\rangle_{a}$ branch, the effective operator is

\begin{equation}
{\langle 0|}_{a_{x}}\ U_{propa}^{x}\ {|0\rangle}_{a_{x}}\  = \ (1\  - \ f_{x})T_{i_{0}}\  + \ f_{x}T_{i_{0} + 1}
\tag{37}
\end{equation}

which is exactly the one-axis restriction of the weighted streaming operator in Eq. (34). The same construction is applied to the y register using an independent ancilla and angle $\theta_{x,y}$. Because the two axes use separate ancillas, the joint postselected operator factorizes as

\begin{equation}
\langle{00|}_{a_{x},\ a_{y}}\ U_{propa}^{x,y}\ {|00\rangle}_{a_{x},\ a_{y}}\  = \ \lbrack(1\  - \ f_{x})T_{i_{0}}\  + \ f_{x}T_{i_{0} + 1}\rbrack\  \otimes \ \lbrack(1\  - \ f_{y})T_{j_{0}}\  + \ f_{y}T_{j_{0} + 1}\rbrack
\tag{38}
\end{equation}

reproducing all four weights in Eq. (34). Streaming acts only within the live radiative subspace $|0\rangle_{sl}$. After streaming, the quantum state is given by

\begin{equation}
\text{|}\phi_{4}\rangle = \sum_{d_{out}}^{}{|d_{out}\rangle\langle}d_{out}| \otimes U_{propa}^{x,y}|\phi _{3}\rangle
\tag{39}
\end{equation}

where all lattice shifts are taken modulo. This construction generalizes straightforwardly to three dimensions by inclusion of an additional $z$-register and corresponding directional control.

\subsubsection{Implementation of the Boundary Condition}

In the classical implementation, absorbing boundary conditions are straightforward. Photons that stream beyond the grid edge are simply discarded using a zero-fill shift instead of a periodic wrap. However, quantum circuits must remain unitary throughout execution, which means amplitude cannot simply be destroyed at the boundary. To resolve this, we introduce two ancilla qubits, ${|escaped\rangle}_{x}$ and ${|escaped\rangle}_{y}$, both initialised to $|0\rangle$, which serve as absorbing boundary markers. During the streaming step, instead of applying a modular increment or decrement that wraps the position register around the grid, we first check whether the photon is attempting to cross a boundary. Specifically, whether the x-register encodes the value $N_{x} - 1$ for a rightward shift (or 0 for a leftward shift), using a multi-controlled X gate targeting the escaped ancilla. If the boundary condition is triggered, the ancilla is flipped to $|1\rangle$ and the position register is left unchanged; otherwise, the standard controlled increment or decrement is applied with the escaped qubit used as an additional control to prevent double-counting. This design is fully unitary because the amplitude is not destroyed but redirected into a subspace labelled by $\left| escaped \right\rangle = |1\rangle$. At the end of each step, during decoding, only amplitudes in the $\left| escaped \right\rangle = |0\rangle$ subspace is read back and mapped to the classical field tensor and amplitudes in the escaped=$|1\rangle$ subspace is discarded, effectively implementing photon absorption at the boundary. The escaped ancilla qubits are reset to $|0\rangle$ before re-encoding for the next step, which is the only non-unitary operation in the entire pipeline and corresponds exactly to the physical irreversibility of boundary absorption. This approach preserves the quantum-classical equivalence

\begin{figure}[htbp]
\centering
\includegraphics[width=5.38542in,height=2.29075in]{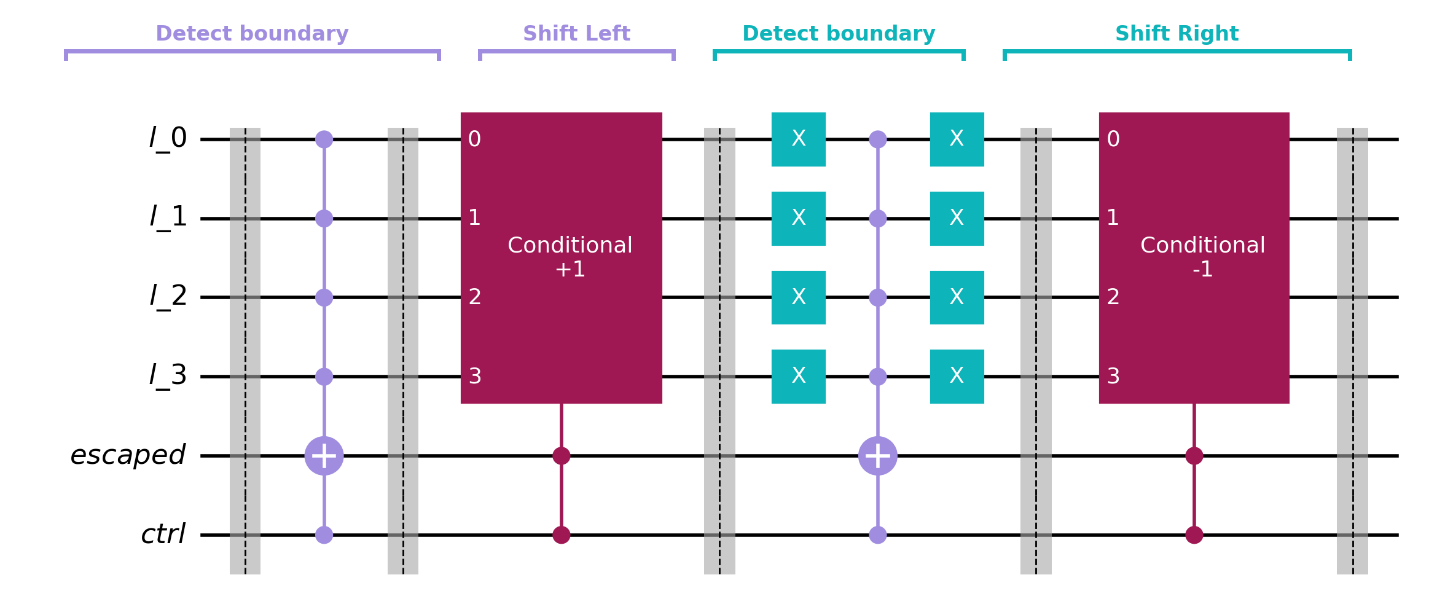}
\caption{Fig. 8: Visualization of the propagation (streaming) gate with boundary conditions: Two detector boundary layers are added for absorbing boundaries; for recycling boundaries, these two layers are discarded.}
\end{figure}

\subsubsection{Decode of the Measurements}

After application of this sequence of quantum update operators, which includes the unitary dilations with ancilla qubits to implement non-unitary scattering, mixing maps of source, extinction through external absorption, and transport operation, the updated radiative field is recovered via post-selection on the ancilla subsystem, and we measure the statistical results from these quantum processes. During quantum calculations, \textbar{}$\phi_{4}\rangle$ denotes the final joint state of the data and ancilla registers from Eq. 35. The final data-register vector ($\phi_{out}$) is obtained by projecting the ancilla onto the designated success state $|0\rangle_{anc}$, hence $\phi_{out} = \langle 0|_{anc}\text{|}\phi_{4}\rangle$. The corresponding post-selection success probability is just the sum of the squared magnitudes of the entries of $\phi_{out}$ ($\ p_{succ} = \parallel \phi_{out} \parallel$, L2 norm). The square root of $p_{succ}$, is a linear estimate of the updated field constructed as $\phi_{out} = \alpha \parallel \phi_{0} \parallel \phi_{in}$ where $\parallel \phi_{0} \parallel$ is the L2 normalization factor used in the initial amplitude encoding of the input field (Eq. 5). $\alpha = \left\| O_{emis} \right\| \times \left\| O_{scat} \right\| \times \left\| O_{abs} \right\| \times \left\| O_{propa} \right\|$ (Eq. 5), and $\alpha$ is the known scaling factor introduced by the dilation procedure. This choice preserves linearity of the update and ensures consistency with the underlying classical transport operator. The recovered vector $v_{out}$is then reshaped into the physical field representation $\ S({s_{st},\ s}_{sl},d_{out},y,x),$ using the same inverse index mapping employed during the encoding step. After post-processing and decoding, the quantum output states are converted into classical physical values.

\section{Results}

Having introduced the complete quantum algorithm and implemented it on the Qiskit Aer simulator, the next step is to validate its correctness. We first validate the proposed 2D algorithm using a 1D configuration, consisting of two propagation directions and a single spatial dimension (Ny=1). The results, presented in Section 3.1, are compared with those of the corresponding 1D algorithm. This comparison verifies that the 2D framework preserves the physical behavior and that the absorption, scattering (without polarization), and emission gates are implemented correctly. We then extend the validation to full 2D simulations by comparing the results obtained from the classical and quantum simulators, as presented in Section 3.2. This further confirms the preservation for the physics correctness of the complete 2D algorithm, demonstrates that the proposed quantum implementation accurately reproduces the expected radiative transfer behavior, and verifies its ability to scale up to higher-dimensional and larger-scale problems.

\subsection{Comparison of the Results from 1D Radiative Transfer Simulations}

For our one-dimensional D1Q2 model without depolarization, both the classical LBM and the proposed QLBM reproduce the solutions of Igarashi et al. (2024) when configured with the same D1Q2 stencil, extinction and backscattering coefficients, as well as identical source and recycle boundary. The classical and quantum solvers yield indistinguishable intensity profiles and attenuation behavior because we are using the simulator that does not have noise problems. We note, however, that Igarashi et al. employed a fused first-order Taylor approximation for extinction, whereas in this study we adopt pure Beer--Lambert exponential extinction without approximation. Under the Beer-Lambert formulation, the peak intensity decreases from 1.806, as reported in their paper, to approximately 1.475 (Figure 9).

To verify implementation consistency between the 1D and 2D solvers, we configured the 2D lattice to emulate a 1D setup in which light propagates exclusively along the west--east direction of the D2Q9 stencil. Under these identical physical and numerical conditions, both the classical and quantum 2D simulations reproduce the corresponding 1D intensity profiles at multiple time steps (Figure 9).

The agreement shown in Figure 9 spans both spatial dimensionality (1D vs 2D) and solver type (classical vs quantum). It confirms that the proposed 2D quantum formulation faithfully reproduces the classical stream--collide dynamics, despite being implemented through direction-controlled lattice shifts and local direction-mixing operators rather than explicit finite-difference updates. Discrepancies between the 2D classical and quantum solvers (on the order of $10^{-12}$) are basically within numerical calculation uncertainties, while differences between the 1D and 2D implementations are even smaller (on the order of $10^{-15}$), regardless of whether the Taylor approximation or Beer--Lambert extinction is used. This validation against the D1Q2 results demonstrates that the current 2D streaming, scattering, absorption, and emission gates correctly reproduce the reference solutions. The corresponding discrepancies for the 2D (D2Q9) quantum simulations are of similar order ($10^{-12}$), albeit with increased circuit depth and size relative to the 1D (D1Q2) implementation.

In the following discussions, we will focus on fully two-dimensional results that incorporate polarization gates and extend the formulation to the D2Q9 lattice.

\includegraphics[width=5.40799in,height=2.65625in]{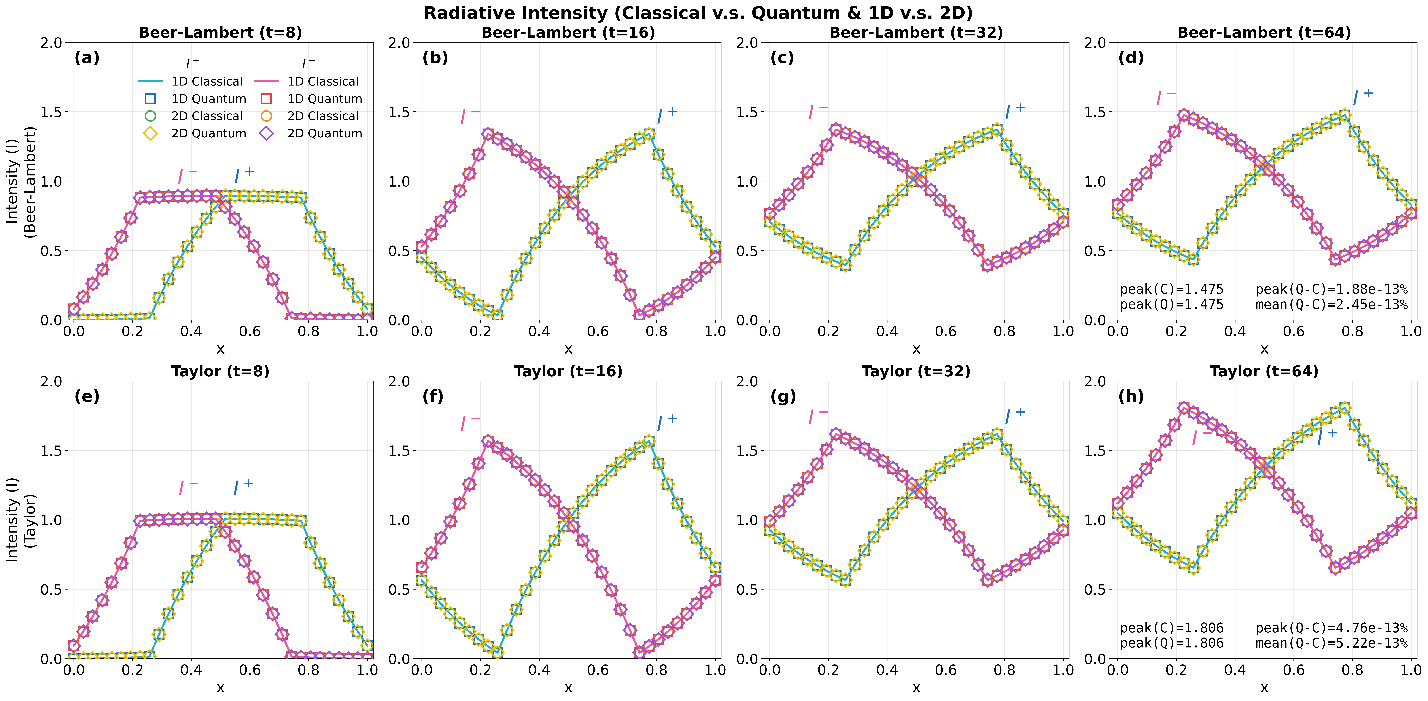}

\textbf{Figure 9}. Reproduction of the D1Q2 benchmark of Igarashi et al. (2024) (I, c1d: classical D1Q2 model; q1d: quantum D1Q2 model) compared with our D2Q9 model (I, c2d: classical D2Q9 model; q2d: quantum D2Q9 model). Quantum intensity fields, computed with Stokes vector representation using the IBM Qiskit simulator, are implemented through 2D quantum circuits. Results are shown for time steps t = 8 (a, e), t = 16 (b, f), t = 32 (c, g), and t = 64 (d, h). The first row (a--d) uses Beer--Lambert extinction; the second row (e--h) uses the Taylor-approximation extinction.

\subsection{Results of 2D Radiative Transfer Simulations}

\subsubsection{QLBM simulation results for D2Q9}

For the full 2D QLBM simulations, we employed a modified model setup incorporating two principal changes. First, the source term is a Gaussian distribution confined to four lattice sites at the center of the domain. This localized, initially symmetric source allows the radiative field to propagate over longer distances, facilitating clearer diagnosis of spatial and directional transport behavior. Second, reduced extinction allows photons to travel farther and enhances the occurrence of multiple scattering events, thereby enabling investigation of depolarization behavior. The core simulation uses molecular Rayleigh scattering together with Beer--Lambert extinction. Owing to memory limitations, we restricted the angular discretization to 8 directions (2 zenith and 4 azimuth angles), yielding a total of 29 qubits. This configuration also simplifies code validation, since photons propagate only along the up/down and east/west/north/south directions. We note that increasing the spatial and angular resolution would make the formulation scalable to larger quantum computers.

Secondly, we consider two different polarization initializations for the source. The first one is [I, Q, U, V] = [1, 0, 0, 0] representing unpolarized (natural) light. The second initiation Stokes vector is equal to [1, 1, 0, 0], representing linearly polarized light aligned with the chosen reference axis. The corresponding intensity and polarization (DOP, Eq. 36) diagnostics are shown below (e.g. Figure 10).

$DOP = \frac{\sqrt{Q^{2} + U^{2} + V^{2}}}{I}$ (Eq. 36)

In Figure 10, the top row shows the two-dimensional map of intensity (I) averaged over all directions, we clearly see that I propagates into four directions as it should as we used only four azimuth angles and with exponent decreasing over the distance to the source. Both quantum and classical results agree very well with the differences on the order of 10-12. The classical simulation values are slightly larger.

The bottom row of Figure 10 shows the polarization (DOP) map. The DOP pattern also behaves as expected: polarization increases with distance as multiple scattering accumulates. For single Rayleigh scattering, polarization for forward and backward directions is zero while it can reach 1 at 90 degrees of scattering angle. At the four edges, scattering saturates around 0.5, while at the corner, the values can pass 0.5 owing to additional blending from adjacent edges and weaker intensity. Overall, the differences in the amplitudes of polarization are on order of $10^{- 12}$, which is also reasonable.

\includegraphics[width=5.37447in,height=3.0625in]{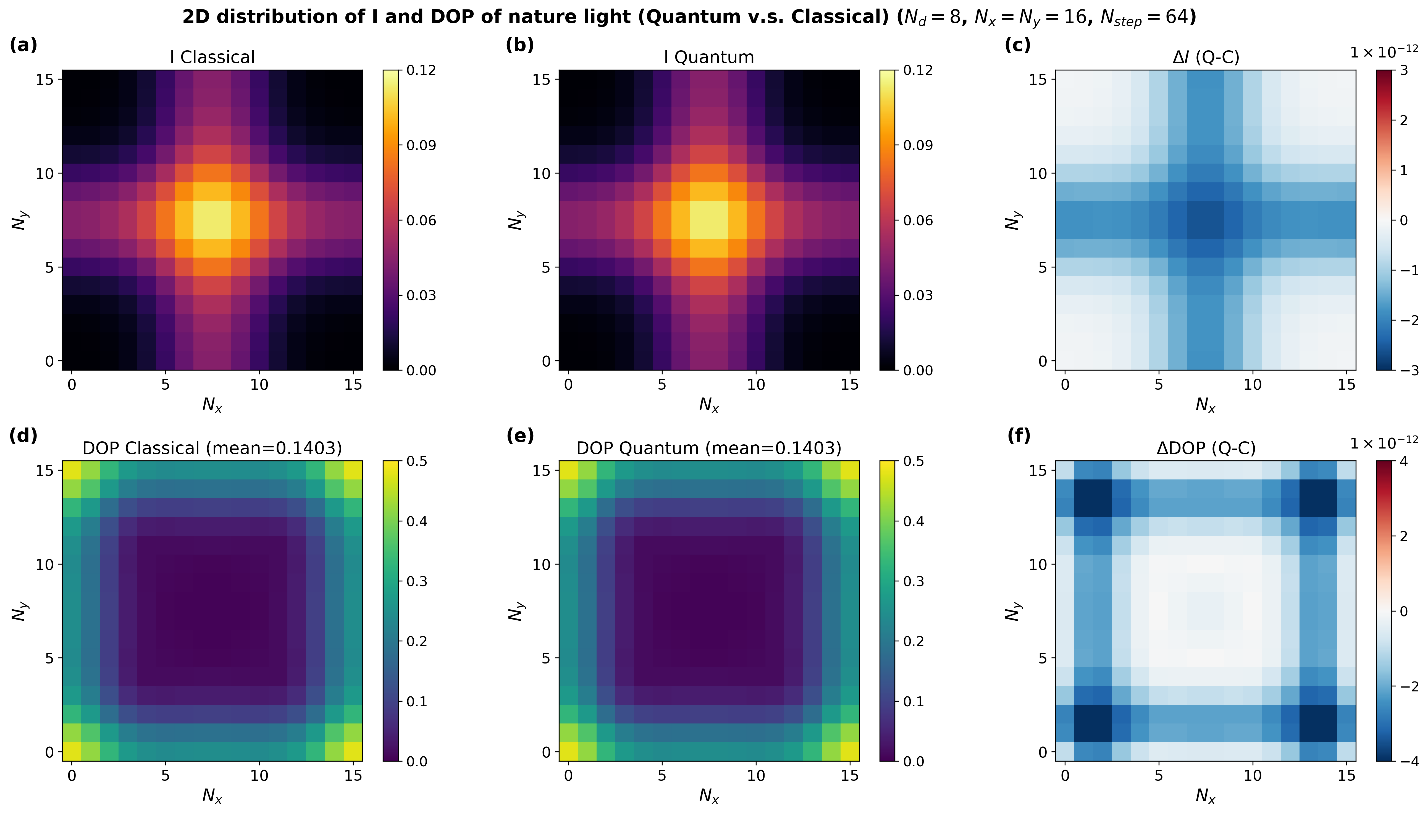} Fig. 10: For natural (unpolarized) light, the intensity (top row) and polarization (bottom row) averaged over 8 directions: classical (a, d) and quantum (b, e) simulation results for the D2Q9 model on a 16$\times$ 16 lattice. Intensity I (a, b) and polarization POL (d, e), with their differences shown in the last column (c, f).

For a source of linearly polarized light, the intensity and polarization spatial maps are shown in Figure 11. The differences between quantum and classical results remain small, on the order of $10^{-12}$. The spatial distribution and transfer pattern of intensity are similar to those of natural light. The main difference compared to natural light is that the polarization decreases from approximately 1 near the source toward approximately 0.5 as the light propagates away from the source.

\begin{figure}[htbp]
\centering
\includegraphics[width=5.539in,height=3.15625in]{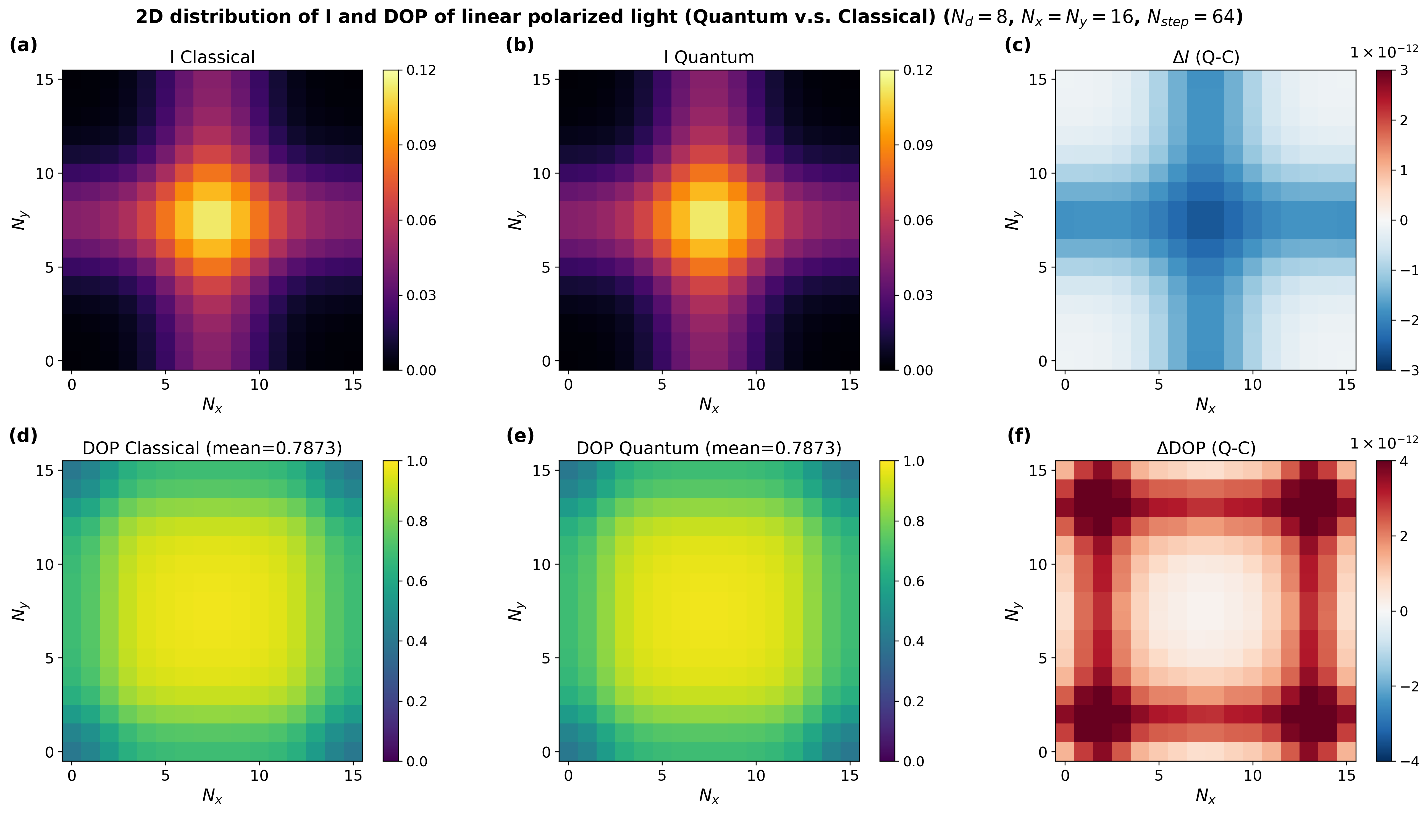}
\caption{Fig. 11: Same as Figure 10, but for a linearly polarized light source.}
\end{figure}

For the absorbing boundary, the quantum and classical results agree to order $10^{-}$12 in both intensity and degree of polarization, confirming that the quantum circuit reproduces the classical transport operator under a bounded domain as well as under periodic wrapping.

The intensity difference between recycling and absorbing boundaries is concentrated on the central cross because at Nd = 8 the angular grid (Naz = 4, Nzen = 2) yields eight streaming directions all aligned with $\pm$ x or $\pm$ y, with no diagonal lattice shifts. Photons therefore leave and re-enter the domain along the same row or column, and since the source at (7, 7) makes row y = 7 and column x = 7 the brightest escape paths, the recycled surplus is deposited there. Measured mean $\Delta$ I is 1.79 $\times$  $10^{-2}$ on the cross against 8.61 $\times$  $10^{-3}$ off it, while the diagonal shows no enhancement (7.62 $\times$  $10^{-3}$). The polarization difference shows the opposite spatial pattern because the edge midpoints lie on the cross and receive the wrapped, nearly unscattered direct beam, which is unpolarized for natural-light injection and therefore dilutes the local DOP, whereas the corners lie off the cross and can only be reached after scattering. Under a Rayleigh phase function, those surviving photons are preferentially scattered near 90$^\circ$, where the induced polarization is maximal, so recycling raises the corner DOP by up to +0.105 while lowering it by $\approx$  -0.059 at the edge midpoints. The two effects nearly cancel in the mean (0.1425 to 0.1403), indicating that the boundary condition redistributes polarization rather than creating or destroying it.

\begin{figure}[htbp]
\centering
\includegraphics[width=5.425in,height=2.26042in]{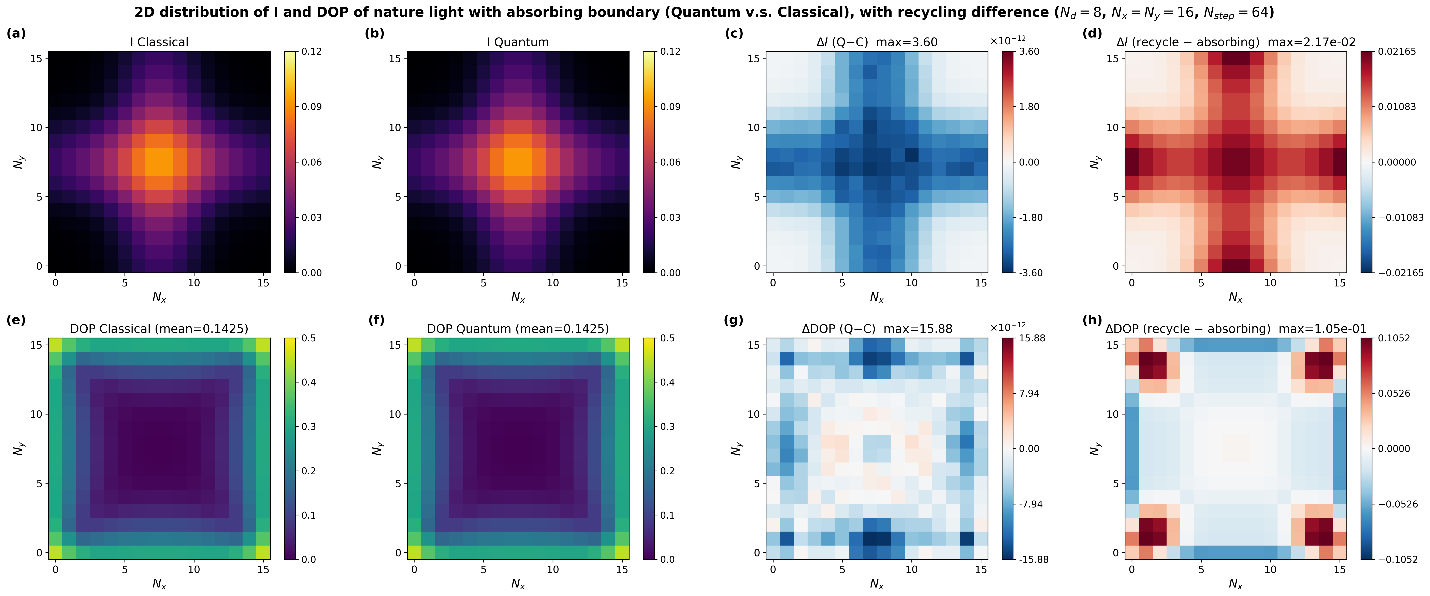}
\caption{Fig. 12: Same as Figure 10, but for an absorbing boundary.}
\end{figure}

For the following discussions and figures, natural lights with recycle boundary only are considered. Figure 13 shows the angular distribution of intensity and polarization as a function of scattering angle and distance. For polarization, the values closest to zero occur in the forward and backward directions, while sideways scattering produces the strongest polarization. This follows the single-scattering Rayleigh polarization behavior (black line in Figure 13i). When separating scattering angles, the polarization stays constant with distance to the source because the DOP at a given scattering angle is set by the Rayleigh phase matrix at the moment of scattering. The intensity, in contrast, is concentrated in the forward direction. This is because the scattering coefficient is weak: most light passes through the molecules and retains its forward direction. The surviving (unscattered) component carries approximately 97\% of the total energy, while the scattering redistribution accounts for only about 3\%. The intensity decays exponentially with distance. The angular differences against scattering angles between quantum and classical results are also small on the order of $10^{- 11}$.

\begin{figure}[htbp]
\centering
\includegraphics[width=5.423in,height=4.45833in]{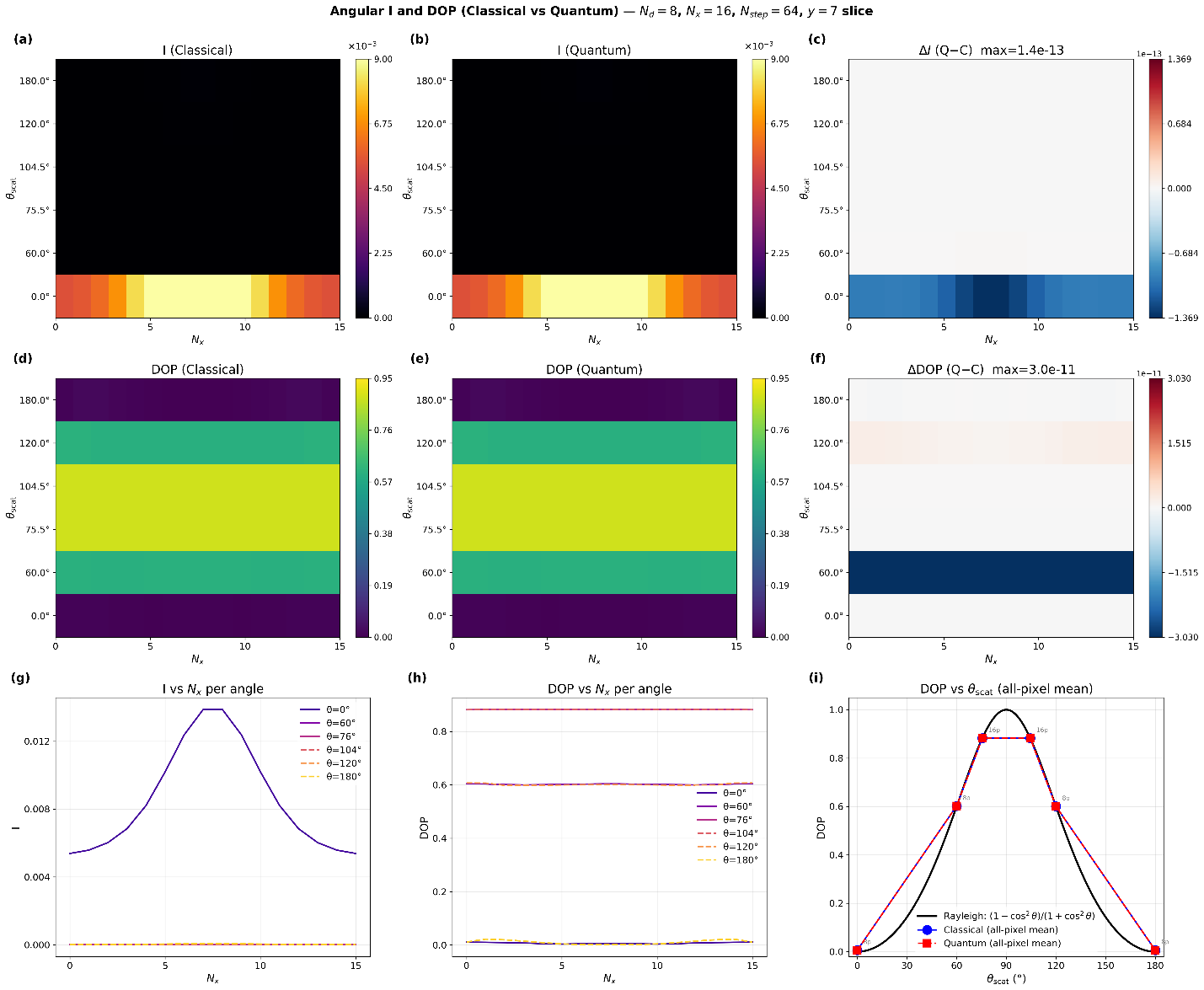}
\caption{Fig. 13: For natural light, the intensity and polarization as a function of scattering angle: classical (a, d) and quantum LBM (b, e) simulation results for the D2Q9 model on an 8$\times$ 8 lattice. Intensity I (a, b) and polarization POL (d, e), with their differences shown in the last column (c, f).}
\end{figure}

\subsubsection{LBM Simulation Results at different Time Step}

We also evaluate the scattering light propagation with the time evolution of the radiative transfer simulation. Figure 14 shows the same quantities as Figure 10, but at time steps 1, 8, 16, and 32 (time step is equal to minimum (1/Nx, 1/Ny), where N is the lattice number). Figure 15 shows the total intensity I, polarization Q, and degree of polarization (DOP), along with the classical--quantum ratios for each, as a function of time. Both classical and quantum radiative transfer simulated lights follow the same light propagation pathway, and the differences remain on the order of $10^{-12}$ across all 64 time-steps, showing no accumulation over time. This stability is an important indication that quantum radiative transfer simulation is temporally scalable.

\begin{figure}[htbp]
\centering
\includegraphics[width=5.41409in,height=6.23836in]{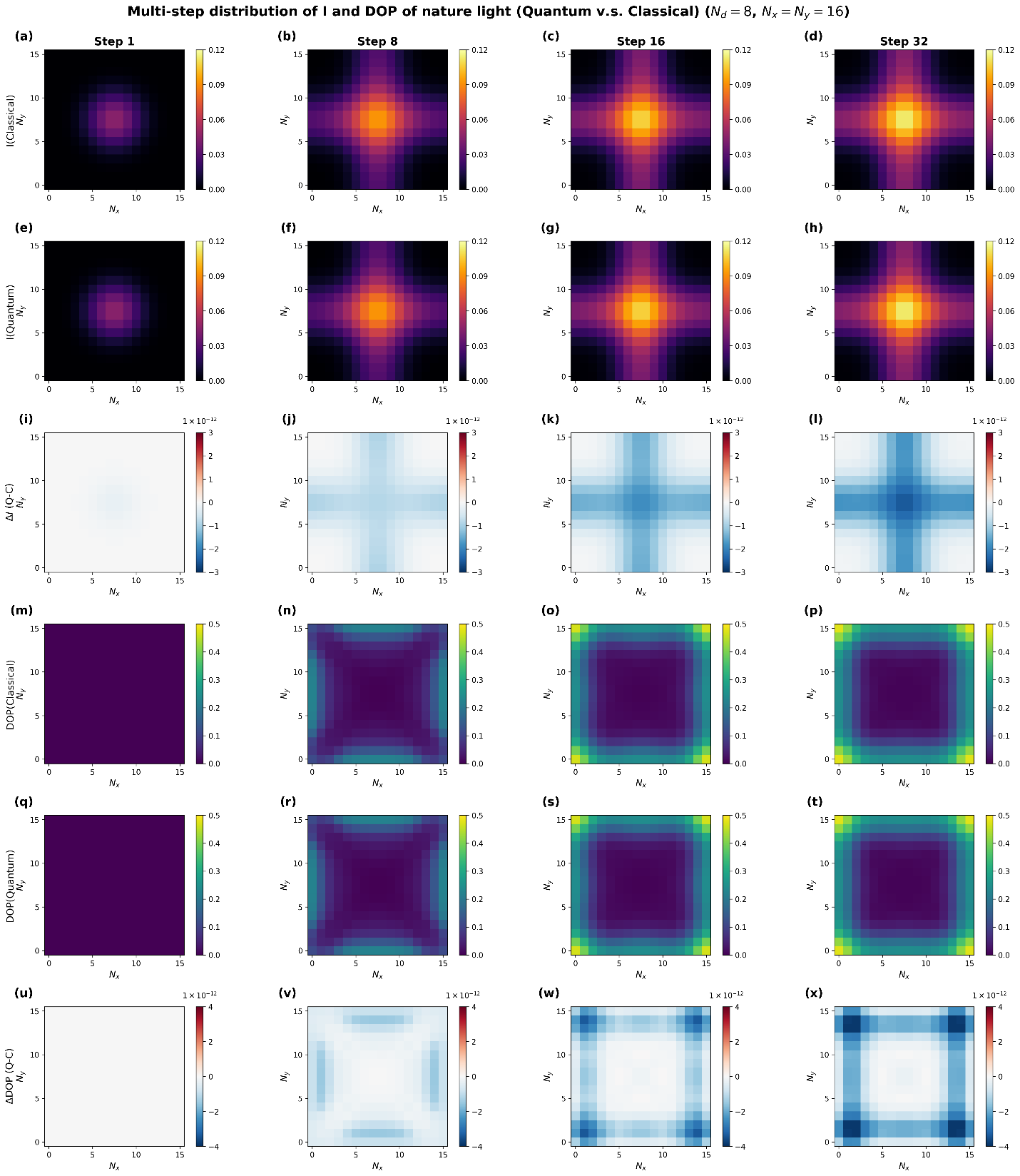}
\caption{Fig. 14: Time series of classical (a to d and m to p) and quantum (e to h and q to t) simulation results for D2Q9 model: I (i to h) and Degree of Polarization, DOP (q to t). Their differences are shown in the third and last rows (I to l for I and u to x for DOP)}
\end{figure}

\begin{figure}[htbp]
\centering
\includegraphics[width=5.42955in,height=3.06736in]{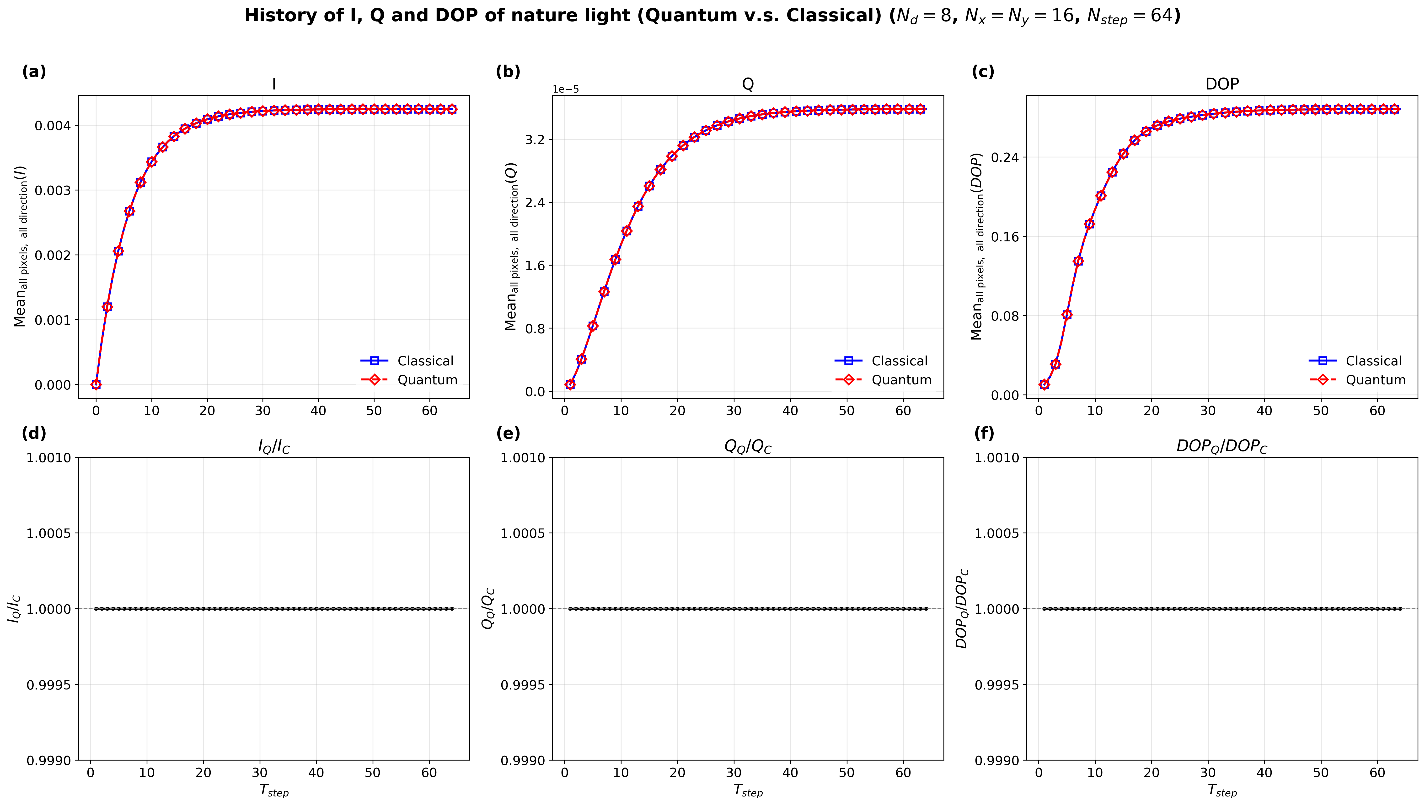}
\caption{Fig. 15: Total I (a), Q (b), and average DOP (c) as a function of time for classical (blue square) and quantum (red diamond) simulations. The corresponding ratios (quantum/classical) are shown in subplots d to f.}
\end{figure}

Figure 16 shows similar results to Figure 15 but for different scattering angles. The polarization associated with each scattering angle is set by the Rayleigh phase function and does not change with time: at step 1 the values are exactly 0.6000 at 60$^\circ$ and 120$^\circ$, and 9/17 = 0.5294 at 75.5$^\circ$ and 104.5$^\circ$, and over 64 steps they changed by less than 0.6\% relative.

The quantum--classical difference shows no angular bias. The relative error is almost the same across all eight propagation directions to within $10^{-11}$; it grows slowly with step count, as expected from accumulated floating-point round-off over successive circuit executions, but does so uniformly rather than preferentially in any direction, as expected. Together with the spatial maps of Figures 10 and 14, this establishes that the quantum and classical simulations agree to within numerical precision in all spatial, temporal and directional domains of degrees of freedom.

\begin{figure}[htbp]
\centering
\includegraphics[width=5.39853in,height=2.99514in]{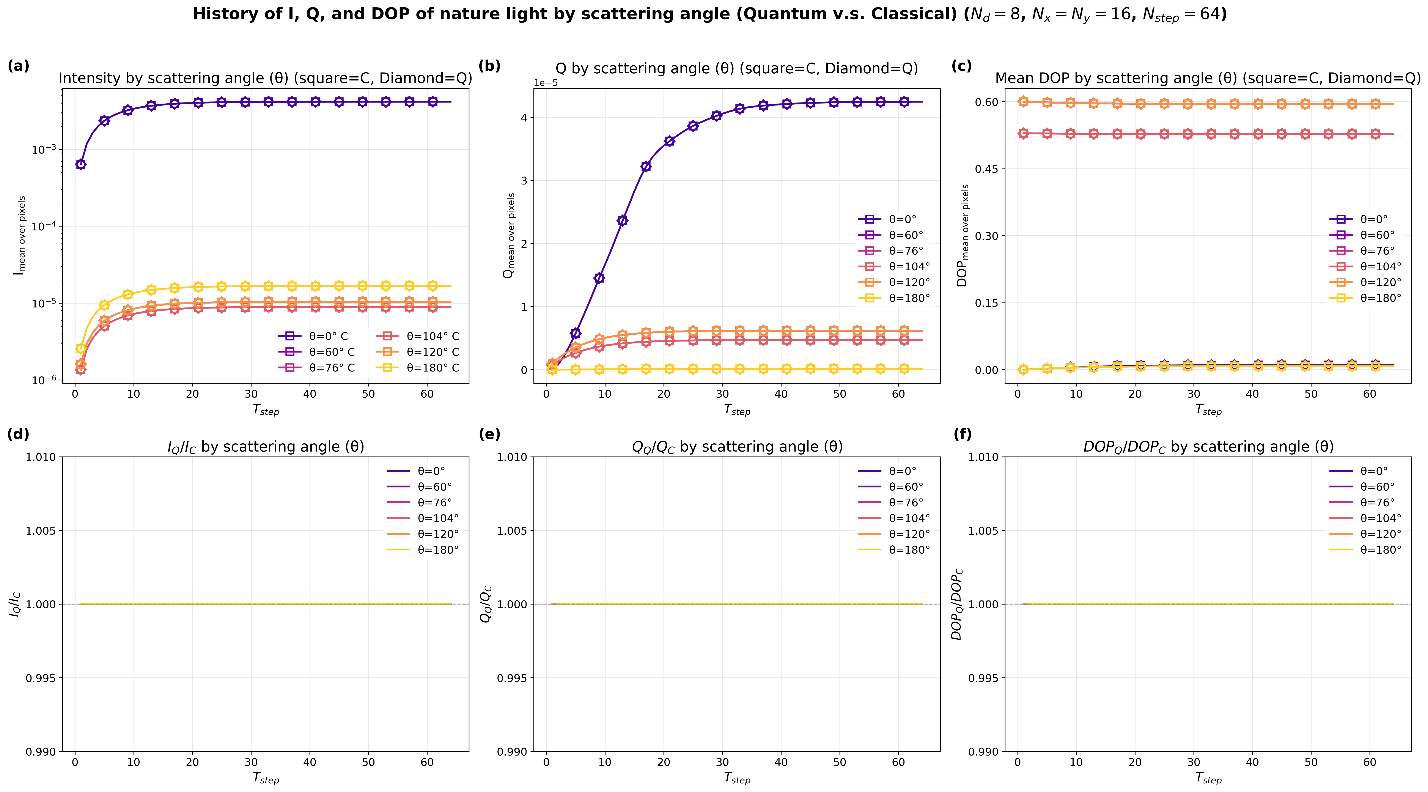}
\caption{Fig. 16: Same as Figure 14 but for different scattering angle.}
\end{figure}

\subsubsection{LBM simulation results Over Multiple Angles Setups}

Because in state vector simulations memory grows exponentially with qubit count, exhaustive testing across high angular resolutions in the simulator is computationally infeasible. For a 2D setup with $N_{x} = N_{y} = 16$ and $N_{d}$ = 8, the simulation encodes approximately 30 qubits, requiring around 4 GB of memory; adding a single qubit doubles this requirement, while scaling both the incoming and outgoing scattering angular dimensions simultaneously results in a fourfold increase in memory consumption. Given these computational constraints, we instead evaluate the correctness of the proposed method across a representative set of zenith and azimuth angle configurations at resolutions that remain computationally tractable.

\textbf{Table 1}. Quantum--classical simulation differences for different azimuth configurations.

\begin{longtable}[]{@{}
  >{\raggedright\arraybackslash}p{(\columnwidth - 10\tabcolsep) * \real{0.1717}}
  >{\raggedright\arraybackslash}p{(\columnwidth - 10\tabcolsep) * \real{0.3348}}
  >{\raggedright\arraybackslash}p{(\columnwidth - 10\tabcolsep) * \real{0.1489}}
  >{\raggedright\arraybackslash}p{(\columnwidth - 10\tabcolsep) * \real{0.1334}}
  >{\raggedright\arraybackslash}p{(\columnwidth - 10\tabcolsep) * \real{0.1187}}
  >{\raggedright\arraybackslash}p{(\columnwidth - 10\tabcolsep) * \real{0.0926}}@{}}
\toprule\noalign{}
\begin{minipage}[b]{\linewidth}\raggedright
\textbf{Configuration}
\end{minipage} & \begin{minipage}[b]{\linewidth}\raggedright
\textbf{Azimuths}
\end{minipage} & \begin{minipage}[b]{\linewidth}\raggedright
\textbf{Polarization}
\end{minipage} & \begin{minipage}[b]{\linewidth}\raggedright
\textbf{I step 4}
\end{minipage} & \begin{minipage}[b]{\linewidth}\raggedright
\textbf{POL step 4}
\end{minipage} & \begin{minipage}[b]{\linewidth}\raggedright
\textbf{L2}
\end{minipage} \\
\midrule\noalign{}
\endhead
\bottomrule\noalign{}
\endlastfoot
Naz=4 cardinal & 0$^\circ$, 90$^\circ$, 180$^\circ$, 270$^\circ$ & natural & 1.9$\times$ $10^{-13}$ & 1.0$\times$ $10^{-14}$ & 1.5$\times$ $10^{-13}$ \\
Naz=4 cardinal & 0$^\circ$, 90$^\circ$, 180$^\circ$, 270$^\circ$ & linear & 1.8$\times$ $10^{-13}$ & 1.7$\times$ $10^{-14}$ & 1.5$\times$ $10^{-13}$ \\
Naz=4 offset & 22.5$^\circ$, 112.5$^\circ$, 202.5$^\circ$, 292.5$^\circ$ & natural & 1.9$\times$ $10^{-13}$ & 7.1$\times$ $10^{-15}$ & 1.5$\times$ $10^{-13}$ \\
Naz=4 offset & 22.5$^\circ$, 112.5$^\circ$, 202.5$^\circ$, 292.5$^\circ$ & linear & 1.8$\times$ $10^{-13}$ & 1.9$\times$ $10^{-14}$ & 1.5$\times$ $10^{-13}$ \\
Naz=3 & 0$^\circ$, 120$^\circ$, 240$^\circ$ & natural & 1.3$\times$ $10^{-13}$ & 8.4$\times$ $10^{-14}$ & 1.0$\times$ $10^{-13}$ \\
Naz=3 & 0$^\circ$, 120$^\circ$, 240$^\circ$ & linear & 1.2$\times$ $10^{-13}$ & 1.6$\times$ $10^{-14}$ & 9.9$\times$ $10^{-14}$ \\
\end{longtable}

\textbf{Table 2}. Quantum--classical simulation differences for different zenith configurations.

\begin{longtable}[]{@{}
  >{\raggedright\arraybackslash}p{(\columnwidth - 12\tabcolsep) * \real{0.0962}}
  >{\raggedright\arraybackslash}p{(\columnwidth - 12\tabcolsep) * \real{0.1603}}
  >{\raggedright\arraybackslash}p{(\columnwidth - 12\tabcolsep) * \real{0.1283}}
  >{\raggedright\arraybackslash}p{(\columnwidth - 12\tabcolsep) * \real{0.1604}}
  >{\raggedright\arraybackslash}p{(\columnwidth - 12\tabcolsep) * \real{0.1283}}
  >{\raggedright\arraybackslash}p{(\columnwidth - 12\tabcolsep) * \real{0.1443}}
  >{\raggedright\arraybackslash}p{(\columnwidth - 12\tabcolsep) * \real{0.1823}}@{}}
\toprule\noalign{}
\begin{minipage}[b]{\linewidth}\raggedright
\textbf{Zenith}
\end{minipage} & \begin{minipage}[b]{\linewidth}\raggedright
\textbf{Polarization}
\end{minipage} & \begin{minipage}[b]{\linewidth}\raggedright
\textbf{I step 1}
\end{minipage} & \begin{minipage}[b]{\linewidth}\raggedright
\textbf{I step 4}
\end{minipage} & \begin{minipage}[b]{\linewidth}\raggedright
\textbf{POL step 1}
\end{minipage} & \begin{minipage}[b]{\linewidth}\raggedright
\textbf{POL step 4}
\end{minipage} & \begin{minipage}[b]{\linewidth}\raggedright
\textbf{L2 step 4}
\end{minipage} \\
\midrule\noalign{}
\endhead
\bottomrule\noalign{}
\endlastfoot
30$^\circ$ & natural & 1.2$\times$ $10^{-13}$ & 1.8$\times$ $10^{-13}$ & 7.5$\times$ $10^{-13}$ & 2.4$\times$ $10^{-13}$ & 1.5$\times$ $10^{-13}$ \\
30$^\circ$ & linear & 1.1$\times$ $10^{-13}$ & 1.6$\times$ $10^{-13}$ & 1.2$\times$ $10^{-14}$ & 1.4$\times$ $10^{-14}$ & 1.4$\times$ $10^{-13}$ \\
45$^\circ$ & natural & 1.2$\times$ $10^{-13}$ & 1.8$\times$ $10^{-13}$ & 1.1$\times$ $10^{-12}$ & 8.4$\times$ $10^{-14}$ & 1.4$\times$ $10^{-13}$ \\
45$^\circ$ & linear & 1.1$\times$ $10^{-13}$ & 1.6$\times$ $10^{-13}$ & 1.4$\times$ $10^{-14}$ & 2.3$\times$ $10^{-14}$ & 1.4$\times$ $10^{-13}$ \\
60$^\circ$ & natural & 1.3$\times$ $10^{-13}$ & 1.9$\times$ $10^{-13}$ & 1.2$\times$ $10^{-12}$ & 1.0$\times$ $10^{-14}$ & 1.6$\times$ $10^{-13}$ \\
60$^\circ$ & linear & 1.2$\times$ $10^{-13}$ & 1.8$\times$ $10^{-13}$ & 1.3$\times$ $10^{-14}$ & 1.9$\times$ $10^{-14}$ & 1.5$\times$ $10^{-13}$ \\
75$^\circ$ & natural & 1.3$\times$ $10^{-13}$ & 2.0$\times$ $10^{-13}$ & 4.3$\times$ $10^{-13}$ & 1.7$\times$ $10^{-14}$ & 1.6$\times$ $10^{-13}$ \\
75$^\circ$ & linear & 1.2$\times$ $10^{-13}$ & 1.9$\times$ $10^{-13}$ & 1.8$\times$ $10^{-14}$ & 2.2$\times$ $10^{-14}$ & 1.6$\times$ $10^{-13}$ \\
\end{longtable}

Table 1 reports the maximum absolute difference between the classical and quantum solutions across several azimuth discretizations --- 3 or 4 directions, both cardinal (0$^\circ$, 90$^\circ$, 180$^\circ$, 270$^\circ$) and offset by 22.5$^\circ$ --- combined with 2 zenith angles. Table 2 reports the same metric across several zenith discretizations --- 4 angles spaced at 15$^\circ$ intervals --- combined with 2 azimuth directions, for both natural and linearly polarized lights.

Across all tested configurations, the quantum implementation agrees with the classical solver to within $10^{-13}$ to $10^{-14}$ in absolute intensity difference, consistent with floating-point round-off at double precision. No systematic deviation was observed as a function of angular offset, polarization mode, or discretization order, confirming that the quantum encoding and scattering gates correctly handle arbitrary direction grids beyond the default axis-aligned case.

\section{Conclusions}

This work presents a lattice-based quantum formulation of atmospheric radiative transfer that addresses multiple scattering, absorption, emission, and polarization through the Stokes vector [I, Q, U, V] presentation. The framework explicitly incorporates polarization processes and extends previous one-dimensional lattice simulations to two spatial dimensions and three directional dimensions, enabling the quantum simulation of coupled transport and polarization dynamics in realistic atmospheric media. This lays the groundwork for fully three-dimensional atmospheric radiative transfer.

Future development will focus on several key directions: spectrally resolved molecular absorption using databases such as HITRAN; aerosol and cloud radiative effects via Mie scattering; and more general scattering representations based on full Mueller matrices. Extending the current two-dimensional D2Q9 lattice formulation to fully three-dimensional lattices remain an important research priority. The present implementation also omits dichroic absorption (polarization-dependent extinction), which will be addressed in subsequence work. This study therefore represents an initial but substantive step toward a physically based quantum radiative transfer solver and, ultimately, a quantum weather and climate modeling framework.

Although the algorithm is designed to be scalable, deployment on fault-tolerant quantum hardware with full error correction remain a longer-term goal. Current quantum devices are constrained by limited qubit counts and high noise levels, preventing practical implementations at this stage. The present formulation operates on a discrete spatial and angular lattice. While this discretization enables efficient numerical implementation and a nature mapping to register-based representations, it inherently limits angular resolution. In particular, large scattering angles are sparsely sampled, which can introduce interpolation artifacts and reduce accuracy in regimes where fine angular structure plays an important role in polarization transport.

In the near term, should the grid size grow prohibitively large, hybrid classical quantum models or machine learning augmented approaches may offer practical alternatives. Looking further ahead, as quantum hardware matures and qubit counts increase substantially, the temporal dimension itself could be encoded in quantum registers, allowing the entire simulation including time evolution to be performed coherently on quantum hardware. This approach, however, is qubit intensive: full unitary time evolution with depolarization tracking requires dedicated ancilla registers to absorb obsolete state values at each time step: fresh "garbage $d_{in}$" registers are required to absorb the old $d_{in}$ values at each step. The total qubit count under this scheme is 27 + 3(N - 1), where N is the number of time steps. For N = 64, this corresponds to 216 qubits, well beyond the capacity of current devices, but a realistic target as hardware capabilities advances.

Last but not least, future work will focus on optimizing circuit designs and simulation frameworks to reduce qubit requirements while improving computational accuracy. In addition, QGU-based chunking for parallel computing and strategies for mitigating hardware noise will be further investigated to enhance scalability, reliability, and simulation efficiency.

\textbf{Data and Software availability.}

All simulations software were performed using the IBM Qiskit software development kit (version 2.3.0, IBM Quantum and Qiskit Community, Qiskit, 2026).

\end{document}